\documentstyle[12pt,psfig]{article}
\begin{document}
\begin{center}
{\Large\bf Representation of a complex Green function on 
a real basis: I. General Theory}
\\[0.5in]

{\bf Robin Shakeshaft}$^a$ {\bf and Bernard Piraux}$^b$\\[0.5in]

$^a$Physics Department\\
University of Southern California\\
Los Angeles, California 90089-0484,USA\\[0.5in]

$^b$Laboratoire de Physique Atomique et Moleculaire\\
2 Chemin du Cyclotron, B-1348 Louvain-la-Neuve, Belgium\\[1.0in]




\def\pref#1{(\ref{#1})}

\def\ie{{i.e.}}
\def\eg{{e.g.}}
\def\cf{{c.f.}}
\def\etal{{et.al.}}
\def\etc{{etc.}}

\def\inbar{\,\vrule height1.5ex width.4pt depth0pt}
\def\IC{\relax\hbox{$\inbar\kern-.3em{\rm C}$}}
\def\IR{\relax{\rm I\kern-.18em R}}
\def\IP{\relax{\rm I\kern-.18em P}}
\def\Z{{\bf Z}}
\def\A{{\bf A}}
\def\B{{\bf B}}
\def\Pone{{\bf P^1}}
\def\gg{{\bf g}}
\def\hh{{\bf h}}
\def\ee{{\bf e}}
\def\aa{{\bf v}}
\def\h{{h_\gg}}
\def\hdual{{h_\gg^\vee}}
\def\jac{{\sl Jac}}
\def\One{{1\hskip -3pt {\rm l}}}
\def\sutwo{{$SU(2)$}}
\def\nth{$n^{\rm th}$}

\def\xtilde{{\tilde X}}
\def\pic{{\sl Pic}}
\def\hom{{\sl Hom}}
\def\prym{{\sl Prym}}

\def\beq{\begin{equation}}
\def\eeq{\end{equation}}

\def\sst{\scriptscriptstyle}
\def\tst#1{{\textstyle #1}}
\def\frac#1#2{{#1\over#2}}
\def\coeff#1#2{{\textstyle{#1\over #2}}}
\def\half{\frac12}
\def\hf{{\textstyle\half}}
\def\ket#1{|#1\rangle}
\def\bra#1{\langle#1|}
\def\vev#1{\langle#1\rangle}
\def\d{\partial}

\def\np{{\it Nucl. Phys. }}
\def\pl{{\it Phys. Lett. }}
\def\pr{{\it Phys. Rev. }}
\def\ap{{\it Ann. Phys., NY }}
\def\prl{{\it Phys. Rev. Lett. }}
\def\mpl{{\it Mod. Phys. Lett. }}
\def\cmp{{\it Comm. Math. Phys. }}
\def\grg{{\it Gen. Rel. and Grav. }}
\def\cqg{{\it Class. Quant. Grav. }}
\def\ijmp{{\it Int. J. Mod. Phys. }}
\def\jmp{{\it J. Math. Phys. }}
\def\nextline{\hfil\break}
\catcode`\@=11
\def\slash#1{\mathord{\mathpalette\c@ncel{#1}}}
\overfullrule=0pt
\def\AA{{\cal A}}
\def\BB{{\cal B}}
\def\CC{{\cal C}}
\def\DD{{\cal D}}
\def\EE{{\cal E}}
\def\FF{{\cal F}}
\def\GG{{\cal G}}
\def\HH{{\cal H}}
\def\II{{\cal I}}
\def\JJ{{\cal J}}
\def\KK{{\cal K}}
\def\LL{{\cal L}}
\def\MM{{\cal M}}
\def\NN{{\cal N}}
\def\OO{{\cal O}}
\def\PP{{\cal P}}
\def\QQ{{\cal Q}}
\def\RR{{\cal R}}
\def\SS{{\cal S}}
\def\TT{{\cal T}}
\def\UU{{\cal U}}
\def\VV{{\cal V}}
\def\WW{{\cal W}}
\def\XX{{\cal X}}
\def\YY{{\cal Y}}
\def\ZZ{{\cal Z}}
\def\lam{\lambda}
\def\eps{\epsilon}
\def\vareps{\varepsilon}
\def\underrel#1\over#2{\mathrel{\mathop{\kern\z@#1}\limits_{#2}}}
\def\lapprox{{\underrel{\scriptstyle<}\over\sim}}
\def\lessapprox{{\buildrel{<}\over{\scriptstyle\sim}}}
\catcode`\@=12

{\bf Abstract}\\
\end{center}

When the Hamiltonian $H$ of a system is represented by a finite 
matrix $\underline{H}$, constructed from a discrete basis with 
overlap matrix $\underline{B}$, the matrix representation 
$\underline{G}(E)\equiv 1/(E\underline {B}-\underline {H})$ 
of the resolvent $G(E)=1/(E-H)$ covers only one branch of $G(E)$. 
We show how all branches can be 
specified by the phase $\phi$ of a complex unit of time 
$t_{\phi}\equiv t_0e^{i\phi}$. This permits $\underline{H}$ to be 
constructed on a real basis; the only duty of the basis is to span 
the dynamical region of space, without regard for the particular 
asymptotic boundary conditions that pertain to the problem of interest. 
Specifically, we show that $P_{\rm cont}G(E)$, where $P_{\rm cont}$ 
projects onto the continuous spectrum of $H$, 
has the series representation
\[
P_{\rm cont}G(E) = P_{\rm cont}e^{-t_{\phi}H}\left(\frac{1}{E}+
2i{t_{\phi}}^2 H\sum_{n=1}^{\infty}\; \frac{1}{n}
{\cal I}_n(Et_{\phi})L_{n-1}^{(1)}(2t_{\phi}H)\right), 
\]
where both the associated Laguerre polynomials, $L_{n-1}^{(1)}(2t_{\phi}H)$, 
and the coefficients ${\cal I}_n(Et_{\phi})$, satisfy 3-term recurrence 
relations.\\[0.5in]

{\large\bf 1 Introduction}\\

The theoretical treatment of a continuous stationary or 
quasistationary process requires some knowledge of the resolvent 
$G(E)\equiv 1/(E-H)$, where $H$ is the Hamiltonian of the system at hand. 
There are numerous methods for evaluating $G(E)$, at least approximately. 
Frequently $H$ is approximated by a finite-dimensional matrix 
$\underline{H}$. The straightforward substitution of $\underline{H}$ 
for $H$ leads to the approximation of $G(E)$ by the matrix $\underline{G}(E)=
1/(E\underline {B}-\underline {H})$ where $\underline {B}$ is 
the overlap matrix for the discrete basis functions used in constructing 
$\underline{H}$. In principle, the basis functions are only 
required to accurately span the (generally 
finite) spatial region where the dynamics take place. However, 
in practice, unless the wavefunction is matched to its asymptotic form 
outside the dynamical region, as is done for example in the R-matrix 
method,\cite{Burk} the basis may also be burdened by the requirement 
that it incorporate the asymptotic boundary condition. This is 
evident from the absence of branch points in $\underline{G}(E)$; the 
expression $1/(E\underline {B}-\underline {H})$ has only poles 
(since the eigenvalue spectrum of $\underline{H}$ is discrete).  
Some of these poles (those in the bound-state region) are legitimate, 
while the others (those in the scattering-state region) simulate 
the branch cut of $G(E)$ by a sequence of discrete points. Hence 
$\underline{G}(E)$ represents, at best, 
only one of the branches of the exact resolvent, and this particular 
branch is determined by the choice of basis. 

In this paper we derive a series representation of the resolvent 
which is such that the asymptotic boundary condition can be imposed 
independently of the basis, without matching the wavefunction; a 
particular branch of $G(E)$ is specified by the phase of a complex 
unit of time. In a companion paper \cite{Sha1} we resum this series to 
give an integral representation of $G(E)$, which is perhaps simpler 
to use in numerical applications. Both the series and integral 
representations can be used even when the Hamiltonian is approximated 
by a real {\em Hermitian} matrix $\underline{H}$, constructed from a 
{\em real} basis. This flexibility may permit a real basis to be used more 
generally and efficiently than otherwise possible, and, in addition, it 
may prove to be useful in the development and practical realization 
of a general theory of continuum 
processes in which detailed knowledge of the asymptotic form of the 
wavefunction is unnecessary. Recently a reformulation of perturbation 
theory for multiphoton processes was given \cite{Sha2} which allows, at 
least in principle, rates and branching ratios to be calculated 
in terms of the flux through a large hypersphere, without input on 
the asymptotic form of the wavefunction. However, the practical 
efficiency of this reformulation depends on the availability 
of a tractable method for representing the resolvent  
without explicit reference to the asymptotic boundary condition.

We pause for a moment to mention some of the difficulties 
that arise when a discrete basis is required to simulate the 
asymptotic behavior of the open channels of a system. For the 
sake of discussion we consider a specific but typical basis 
that is used to construct the Hamiltonian of a one-electron 
system such as a hydrogen atom; the basis functions are
\[
\sqrt{-2i\kappa}(2i\kappa r)^{l+1}e^{i\kappa r}P_n(2i\kappa r)
Y_{lm}({\hat {\bf x}}), 
\]
where $P_n(x)$ is a polynomial of degree $n$, where $Y_{lm}({\hat {\bf x}})$ 
is a spherical harmonic, and where $\kappa$ is a parameter --- the 
{\em wavenumber} of the basis. This basis can be readily generalized 
to a multiparticle system by specifying the orientation of the system 
in terms of Euler angles and by introducing products of radial 
functions, one in each of the interparticle distances (or 
combinations of them), perhaps associating a different 
wavenumber with each distance. A bound-state wavefunction satisfies 
a {\em real} damped-wave asymptotic boundary condition, which 
can be described by choosing a real basis with positive pure 
imaginary wavenumbers. Hence a real basis is well-suited to the description 
of closed channels and to the study of bound-state properties.\cite{Pek} 
However, when $\underline{G}(E)$ is constructed from a real basis it 
has spurious poles on the ``unitarity'' branch  cut of $G(E)$ along the 
positive real energy axis, and these poles give rise to spurious resonances 
with zero width in the scattering-state energy region. The wavefunction 
for a compound (metastable) state satisfies a {\em complex} outgoing-wave 
asymptotic boundary condition, which can be described by choosing a 
complex basis with wavenumbers lying in the upper-right quadrant of the 
wavenumber plane; 
\footnote{Choosing the basis wavenumbers to lie in the upper right quadrant of 
the wavenumber-plane is equivalent to rotating the particle coordinates 
into the lower right quadrant of the complex position-coordinate-plane. 
See e.g. \cite{Rein}.}
since the factor $e^{i\kappa r}$ behaves as a damped outgoing-wave,
this basis is well suited to the 
description of both closed channels {\em and} {\em out}going-wave 
open channels, and can be used to treat half-collision process, e.g. 
autoionization \cite{Ho} or photoionization \cite{Klar}. Furthermore, 
$\underline{G}(E)$ now has 
poles in the scattering-state region that lie in the upper half of the 
complex energy plane, and they simulate a branch cut along a line which 
is distinct from the real axis, so spurious resonances are only of 
minor significance. However, a price is paid for 
choosing the basis to be complex, i.e. $\underline{H}$ is 
{\em non}Hermitian, and therefore unitarity is only approximately 
satisfied. In fact, the norm $\langle \Psi|\Psi \rangle$ of a state 
vector $|\Psi \rangle$ is a nonanalytic function of $\kappa$ since the 
bra $\langle \Psi|$ depends on $\kappa^*$; consequently, the norm does 
not converge as the basis size is increased. In addition, it is 
difficult to extract partial rates, i.e. branching ratios, since 
a partial rate is also nonanalytic in $\kappa$. Finally, a complex 
basis cannot, in general, describe the {\em real standing-wave} boundary 
condition satisfied by the wavefunction for a full-collision 
process. 
\footnote{
An exception occurs when the scattering potential falls off at least as 
fast as an exponential potential. In this case the deviation of 
the wavefunction from a standing wave is inconsequential at large 
distances. A wavefunction that is a real standing wave over short 
distances has a finite number of nodes, and it {\em can} be constructed 
from a complex basis provided that polynomials $P_n(2i\kappa r)$ of 
sufficiently high degree are included. At the same time 
the correct branch of $\underline{G}(E)$ is guaranteed by the 
factor $e^{i\kappa r}$. See e.g. \cite{Baum}. }

While a real basis, with pure imaginary wavenumbers, can describe a 
standing-wave boundary condition, the branch of $G(E)$ cannot be 
specified uniquely by real basis functions since they do not 
distinguish between ingoing- and outgoing-wave behavior. On the 
other hand, when a real basis is used in conjunction with the 
series and integral representations developed here, the only duty of the 
basis is to span the dynamical region of space. The only purpose of 
the purely decaying exponential factor $e^{i\kappa r}$ ($\kappa$ is 
positive pure imaginary) is to restrict the range of the 
basis to the dynamical region.

In order to impose asymptotic boundary conditions that are independent of 
the basis we start from the observation that all the particles in 
a system evolve according to a common time, and the asymptotic 
behaviour of the system's wavefunction emerges from the initial boundary 
conditions, at the start of the system's evolution, say time 
$t=0$. To simplify our analysis we consider a system comprised 
of only one particle; the generalization to more 
than one particle is fairly straightforward and will be dealt with 
elsewhere. The temporal behavior is governed by the time-evolution 
operator $U(t)\equiv e^{-iHt}$ (we set $\hbar=1$ throughout). Let us introduce 
the dimensionless variable $\tau\equiv t/t_0$, where $t_0$ is the 
unit of time that characterizes the time scale on which the 
motion of the system occurs. If $\mbox{Im}\; E>0$ we can represent 
$G(E)$ by an integral along the positive $\tau$-axis: 
\begin{equation}
G(E) = -it_0\int_0^{\infty}d\tau \; e^{i(t_0E)\tau}e^{-i(t_0H)\tau}.
\label{eq:101}
\end{equation}
We can analytically continue this representation of $G(E)$ to a sector 
of the $E$-plane that includes the negative real $E$-axis if we 
first project out the bound states, i.e. if we remove the bound-state 
poles of $G(E)$. Thus we consider $P_{\rm cont}G(E)P_{\rm cont}$, 
where $P_{\rm cont}$ is the projection operator $P_{\rm cont}=1-
P_{\rm bd}$, with $P_{\rm bd}$ the bound-state projection operator. 
We rotate the contour of integration from the positive real 
$\tau$-axis through a small angle into the lower-right quadrant 
of the complex $\tau$-plane, and thereby obtain a representation of 
$P_{\rm cont}G(E)P_{\rm cont}$ which is valid in a region 
including the negative real $E$-axis. We can represent $G(E)$ 
throughout the lower half of the $E$-plane by an integral along the 
negative $\tau$-axis: 
\begin{equation}
G(E) = -it_0\int^{-\infty}_0d\tau \; e^{i(t_0E)\tau}e^{-i(t_0H)\tau},
\label{eq:102}
\end{equation}
and again by considering $P_{\rm cont}G(E)P_{\rm cont}$, and  
rotating the integration contour through a small angle (now into the 
lower-left quadrant of the $\tau$-plane) we obtain a representation of 
$P_{\rm cont}G(E)P_{\rm cont}$ which is valid in a sector of the 
$E$-plane that also includes the negative real $E$-axis. As we let 
$E$ approach the positive real axis these two different representations 
yield two different branches of $P_{\rm cont}G(E)P_{\rm cont}$, and 
since both representations give the same ``physical'' 
branch of $P_{\rm cont}G(E)P_{\rm cont}$ on the negative real $E$-axis 
we can define a global representation of $P_{\rm cont}G(E)P_{\rm cont}$ 
on the first --- the  ``physical'' --- sheet of the Riemann energy 
surface by using Eq.~(\ref{eq:101}) for $\mbox{Im}\; E\geq 0$ 
and Eq.~(\ref{eq:102}) for $\mbox{Im}\; E\leq 0$. The point is: A 
particular branch of $P_{\rm cont}G(E)P_{\rm cont}$ on the 
positive energy axis can be specified by 
the contour of the global integral representation. Our goal is to perform 
the integration over $\tau$ in such a way that a signature of the 
contour is preserved. 

That signature is the angle of rotation of the contour, whose vestige 
is the phase $\phi$ of a complex unit of time 
$t_{\phi}\equiv t_0e^{i\phi}$. The main result of our paper is 
that the resolvent, with the bound-states removed, can be expressed 
generally as the series 
\begin{equation}
P_{\rm cont}G(E) = P_{\rm cont}e^{-t_{\phi}H}\left(\frac{1}{E}+
2i{t_{\phi}}^2 H\sum_{n=1}^{\infty}\; \frac{1}{n}
{\cal I}_n(Et_{\phi})L_{n-1}^{(1)}(2t_{\phi}H)\right), 
\label{eq:121}
\end{equation}
where $L_{n-1}^{(1)}(2t_{\phi}H)$ is an associated Laguerre polynomial of 
degree $n-1$ in the operator $2t_{\phi}H$, and where the coefficient 
${\cal I}_n(Et_{\phi})$ is a number defined by the integral
\begin{equation}
{\cal I}_n(a) \equiv \int_0^{\infty}d\tau \; e^{ia\tau}
\left( \frac{\tau+i}{\tau-i}\right)^n .\label{eq:122}
\end{equation}
The energy, $E$, can take on any value on the first sheet of the 
Riemann surface cut along the line arg~$(E)=-\phi$. The 
whole Riemann surface can be covered by rotating the cut, i.e. by 
varying the phase $\phi$. The magnitude, $t_0$, of the unit of time, 
$t_{\phi}$, is a parameter whose value is to some extent, but not 
entirely, arbitrary. It should not differ greatly from the 
characteristic unit of time for the motion of the 
system, i.e. from the time it takes for the wavefunction of the 
system to change appreciably, for otherwise convergence of the 
series would be difficult to attain. In particular, $t_0$ should 
not exceed the characteristic duration of the process under study, 
since a measurable process cannot take place in a time less 
than it takes for the wavefunction to evolve appreciably. For 
example, the characteristic duration of an elastic collision process 
at energy $E$ is of order $1/E$ so that in this case we require $Et_0<1$. 
For asymptotically large values of $E$ 
the time-scale is infinitesimally short, i.e. $t_{\phi}\sim 0$, and 
only the first term on the right side of Eq.~(\ref{eq:121}) contributes, 
giving $G(E)\sim 1/E$ for all branches. Apart from this $1/E$ term, 
the dependence of $G(E)$ on $E$ is contained in the coefficients 
${\cal I}_n(Et_{\phi})$, and a particular branch of $G(E)$ is fixed 
by these coefficients after specifying the phase $\phi$. 
The physical branch of $G(E)$, which 
has outgoing-wave behavior, is specified by choosing $\phi$ to be in 
the range $0< \phi < \pi$, and the other, ``unphysical'', branch, 
which has ingoing-wave behavior, is specified by choosing $\phi$ 
to be in the range $0> \phi > -\pi$. 
In principle, the right side of Eq.~(\ref{eq:121}) should not vary as 
$t_{\phi}$ varies over its allowed range, but in practice, when 
the series is truncated and the summation not fully converged, 
it does vary --- the less so, the better the convergence. 
The presence of $P_{\rm cont}$ in Eq.~(\ref{eq:121}) is perhaps not 
surprising since we were obliged to project out the bound states before 
writing down a global {\em integral} representation of the resolvent. 
[Actually, since $G(E)$ acts, in general, on a wavepacket, we do not 
need to multiply the Hamiltonian by $P_{\rm cont}$; rather, we need 
only let $P_{\rm cont}$ act on the wavepacket.] When the Hamiltonian 
contains a potential that has an attractive 
Coulomb tail, an infinite number of bound states accumulate at threshold, 
and it is neither necessary nor desirable to remove those bound 
states just below threshold; we elaborate further on this in 
sections 2 and 3.

Note that the Laguerre polynomials satisfy a 3-term recurrence relation, 
and therefore they can be calculated recursively with only a minimal 
number of multiplications of $\underline {H}$. In addition, as 
shown in Appendix C, the coefficients ${\cal I}_n(Et_{\phi})$ can be 
calculated using a 3-term recurrence relation. Furthermore, 
since the evolution operator $e^{-t_{\phi}H}$ acts for only 
one unit of time, it can be calculated using a Pad\'{e} 
approximant --- see Appendix C. 

In the derivation of Eq.~(\ref{eq:121}) which we give in section 4, 
we initially restrict the phase of $t_{\phi}$ to the range 
$-\pi/2 <\phi<\pi/2$. However, the series can be analytically continued 
outside this range. The preliminary restriction on the phase is related 
to a restriction on the contour of the global integral representation,  
as we now explain. For the sake of discussion, assume $E$ to have a 
physically realizeable value, one appropriate to a collision process, 
i.e. $E$ real and positive. 
The integrals of Eqs.~(\ref{eq:101}) and (\ref{eq:102}) differ only 
by their contours. These contours can be rotated from the positive 
and negative real $t$-axes into the upper half of the $t$-plane, 
but they must remain distinct since the integrals represent two 
different branches of $G(E)$. Hence if we choose one contour 
to lie along the ray arg~$(t)=\phi$ in the upper-right quadrant, 
with $\phi$ therefore restricted to $0\leq \phi <\pi/2$, we should 
choose the other contour to lie along the ray arg~$(t)=\pi+\phi$ 
in the upper-left quadrant, with $\phi$ restricted to 
$-\pi/2<\phi\leq 0$; therein lies our preliminary restriction on the 
phase. 

To gain an intuitive understanding of the series, 
we give now a brief {\em heuristic} derivation of Eq.~(\ref{eq:121}). 
We start by formally expressing $G(E)$ as a Taylor series in $H/E$: 
\begin{equation}
G(E) = \frac{1}{E}+\frac{1}{E}\sum_{n=1}^{\infty}\; 
\left( \frac{H}{E} \right)^n .\label{eq:123}
\end{equation}
This series is an {\em asymptotic} series in powers of $H/E$. It is  
applicable for large $E$, i.e. it is applicable on a short time-scale, but 
not in all sectors of the complex $E$-plane. In fact, it is not applicable 
when $E$ is real and positive; for if $H$ is also positive, i.e. if 
$H$ acts on those of its eigenvectors belonging to the continuous 
eigenvalue spectrum --- whose threshold we choose to be 
zero --- {\em all} of the terms $(H/E)^n$ are {\em positive} and 
the sum steadily diverges as more terms are included. Hence, for $E$ 
real and positive we are motivated to delete continuous eigenvalues much 
larger than $E$ by multiplying the right side of Eq.~(\ref{eq:123}) by a 
cutoff factor. A natural choice for  
the cutoff factor is $e^{-t_{\phi}H}$, provided the following 
restrictions are imposed on $t_{\phi}$: In order for $e^{-t_{\phi}H}$ 
to cut off the sum we require that Re~$t_{\phi}>0$, i.e. 
$|\phi| < \pi/2$, but since we do not need (or want) to delete 
continuous eigenvalues much smaller than $E$ we require that $Et_0<1$; 
these are the same restrictions on $t_{\phi}$ mentioned above. 
(We note again that the restriction on $\phi$ can 
be lifted by analytic continuation.) 
We can regather the terms in the sum on the right side of 
Eq.~(\ref{eq:123}), now modified by the cutoff factor, to yield an 
expansion in the basis functions $xe^{-x/2}L_n^{(\nu)}(x)$, 
where $x=2t_{\phi}H$ and where the $L_n^{(\nu)}(x)$ are associated 
Laguerre polynomials. These basis functions satisfy the end-point 
boundary conditions for the modified sum, i.e. they vanish linearly 
at $x=0$ and exponentially at $x=\infty$. Furthermore, these basis 
functions form a complete set over the interval $0\leq x <\infty$, 
and they satisfy the orthogonality relation
\begin{equation}
\int_0^{\infty}dx\; x^{\nu}e^{-x}L_m^{(\nu)}(x)L_n^{(\nu)}(x)
=\frac{\Gamma(\nu+n+1)}{\Gamma(n+1)}\delta_{mn}. \label{eq:124}
\end{equation}
Here $x$ is understood to be a continuous variable, and consequently 
we must restrict the operator $H$ to act only on those 
of its eigenvectors belonging to the continuous eigenvalue spectrum. 
In other words, we must exclude eigenvectors belonging to the 
discrete eigenvalue spectrum. Choosing the upper index $\nu$ of the 
Laguerre polynomials to be 1, it follows that
\begin{equation}
\frac{P_{\rm cont}}{E-H}
=P_{\rm cont}e^{-t_{\phi}H}
\left( \frac{1}{E}+it_{\phi}(2t_{\phi}H)\sum_{n=1}^{\infty}\; 
\frac{1}{n}{\cal I}_n(Et_{\phi})L_{n-1}^{(1)}(2t_{\phi}H)\right) ,
\label{eq:125}
\end{equation}
where the coefficients, determined using Eq.~(\ref{eq:124}), are 
\begin{equation}
{\cal I}_n(Et_{\phi}) = -2i\int_0^{\infty}dH \; e^{-t_{\phi}H}\left(\frac{1}
{E-H}-\frac{e^{-t_{\phi}H}}{E}\right)L_{n-1}^{(1)}(2t_{\phi}H), \label{eq:126}
\end{equation}
where here $H$ is understood to be a nonnegative continuous variable 
of integration. Thereby we arrive at Eq.~(\ref{eq:121}). While not 
obvious, the coefficients defined by Eq.~(\ref{eq:126}) 
are the same as those defined by Eq.~(\ref{eq:122}); 
we prove this in Appendix A. Although Eq.~(\ref{eq:125}) was just derived  
for $E$ real and positive, the results can be analytically continued 
to complex $E$.

It is time to remark on other polynomial expansions of 
the resolvent. If the Hamiltonian is represented by a finite matrix, 
whose minimum and maximum eigenvalues are 
$E_{\rm min}$ and $E_{\rm max}$, respectively, one can, by fiat, 
expand the matrix representation of the resolvent in any 
set of orthogonal polynomials that form a complete 
set over a finite interval rescaled to the interval 
$[E_{\rm min},E_{\rm max}]$. The most popular of these expansions uses 
Chebyshev or Faber polynomials \cite{Kour,Kour2} since the evolution 
operator $U(t)\equiv e^{-iHt}$ can be expanded very effectively over 
a finite time interval in terms of these polynomials.\cite{Kour2,Tal} 
The Chebyshev expansion has been remarkably successful in obtaining 
properties of bound and compound states. \cite{Mand} However, the 
Chebyshev and Faber expansions do not discriminate between the discrete 
and continuum parts of the spectrum of $H$, and notwithstanding some 
numerical evidence that these expansions can yield reasonably accurate 
estimates of scattering amplitudes in some simple cases, \cite{Kour,Kour2} 
it is unclear at this stage how well-suited are these expansions for the 
treatment of general collision processes, particularly when the Coulomb tail 
plays an important role. The long-time behavior of the temporal 
correlation amplitude (see below) is determined by the spectral density at the 
threshold of the continuous spectrum. The success of the Chebyshev 
expansion in dealing with discrete (bound and compound) states 
can be understood by observing that each of these states 
decays exponentially with increasing time in a wide sector of the 
complex time-plane; hence the effective time-interval over which 
$U(t)$ must be expanded is relatively 
short. However, in a continuum process, wavepacket spreading 
occurs, and its asymptotic temporal behavior is $1/t^{3/2}$ 
(for short-range potentials), a rather slow fall-off with $t$. 

Our primary interest is in stationary processes in the continuum, for 
which $E$ is real and positive. We focus in the present paper on the 
general theory underlying the series representation of the resolvent. 
As noted already, for simplicity we consider a system comprised 
of only one particle. Our analysis is largely informal (a 
mathematically rigorous treatment is, anyway, beyond our abilities). 
In a companion paper \cite{Sha1} we derive the integral representation, 
and we give numerical illustrations for the examples of photoionization 
of a hydrogen atom and s-wave scattering from a $1/(1+r)^4$ potential. 
We plan to report on an application to a multiparticle system 
in the future. 

We frame our paper within the context of the inclusive rate at which 
a continuous stationary or quasistationary process occurs. This  
rate can be expressed in terms of a matrix element of $G(E)$ 
of the form $\langle \psi |G(E)|\psi \rangle$ where 
$|\psi \rangle$ is a localized wavepacket. \cite{Gold} This 
matrix element can in turn be expressed, according to Eq.~(\ref{eq:101}), 
as $-i\int_0^{\infty}dt\; e^{iEt}C(t)$ where $C(t)$ is the temporal 
correlation amplitude:
\begin{equation}
C(t)\equiv \langle \psi |\psi (t)\rangle, \label{eq:201}
\end{equation}
with $|\psi (t)\rangle \equiv U(t)|\psi \rangle$. 
In the next section we explore the analytic properties of the correlation 
amplitude in the complex-time-plane. We find that the singularities 
of $C(t)$ lie in the upper half of the $t$-plane. While we can 
expand $U(t)$, and thereby $C(t)$, in powers of $t$, 
the power series for $C(t)$ has only a small (maybe infinitesimal)  
radius of convergence. In section 3 we transform 
variables, from $t$ to a variable $u$, to obtain a power series 
which is more useful. Thus we make a conformal transformation 
which maps a singularity-free region of the $t$-plane --- namely, 
the half of the complex $t$-plane that lies below the line 
$\mbox{Im}\; t/\mbox{Re}\; t = \tan \phi$ --- 
into the unit circle in the $u$-plane. We can express $U(t)$ and 
$C(t)$ as power series in $u$, with coefficients $c_n$ that are rather 
simply related to the coefficients of the power series in $t$. 
We analyse the properties, in particular, the large-$n$ behavior, of 
the coefficients $c_n$, though we defer some of the analysis to 
Appendix D. In section 4 we perform the integration over $t$ and obtain 
the series representation, Eq.~(\ref{eq:121}), of $G(E)$. We analyse 
the convergence properties of this series, and also show that 
the higher terms in the series can be resummed as another series 
that converges rapidly. 
In Appendix A we prove the equivalence of the two forms of 
${\cal I}_n(a)$ introduced above. In Appendix B we use the series 
representation of the resolvent to reproduce some known formal results, 
for example, threshold laws. In Appendix C we describe some algorithms 
that are useful for the implementation of the series, and in Appendix D 
we analyse the large-$n$ behavior of the coefficients $c_n$ when 
the potential has a Coulomb tail.\\

{\large\bf 2 Analytic Properties of the Correlation Amplitude}\\

Recall that $|\psi \rangle$ is a normalizeable, localized wavepacket  
which evolves in time $t$ as $|\psi (t)\rangle =U(t)|\psi \rangle$, 
where $U(t)\equiv e^{-iHt}$ is the time-evolution operator and where 
$H$ is the Hamiltonian of a system that 
is comprised of only one particle, of mass $\mu$ say. We assume that the 
wavepacket $|\psi \rangle$ includes a continuous superposition of 
scattering eigenstates of $H$, and is not merely a superposition of 
discrete (bound- and compound-) eigenstates. If the potential has an 
attractive long-range Coulomb tail, a {\em localized} wavepacket has, in 
general, a nonzero overlap with an infinite number of bound-eigenstates of 
$H$. Indeed, if $L$ is the characteristic linear dimension of $|\psi \rangle$ 
in position space, Rydberg bound states with energy eigenvalues greater 
than or of the order of $-\mu/L$ are indistinguishable in the composition 
of the wavepacket from scattering eigenstates with energy eigenvalues 
less than or of the order of $\mu/L$. As a consequence, 
high Rydberg bound states, through their contribution 
to the correlation function, play an important role in continuum 
processes, a feature we observe below. The case where the potential 
has a Coulomb tail is usually an exception requiring special 
treatment. Leaving aside this 
case for the moment, the continuum portion of $|\psi (t)\rangle$ spreads 
linearly in time, and therefore occupies a volume (in 3-dimensional 
position space) that is proportional to $t^3$. Since  
$U(t)$ is unitary, $\langle \psi (t)|\psi (t)\rangle$
is conserved in time, and so the continuum portion of the wavepacket 
attenuates at each point in space as $t^{-3/2}$. Consequently, 
the correlation amplitude $C(t)\equiv \langle \psi |\psi (t)\rangle$ 
contains a component which vanishes as $t^{-3/2}$ 
for $t\sim \infty$. Therefore $C(t)$ 
has a branch point at infinity, and since this branch point 
is of order two there must be another branch point, 
joined to the branch point at infinity by a cut in the 
complex $t$-plane. 

As an example, consider a free particle (of mass $\mu$) whose 
position is initially described by the Gaussian wavepacket
\begin{equation}
\langle {\bf x}|\psi \rangle 
=(\kappa_0^2/\pi)^{3/4}e^{-\frac{1}{2}\kappa_0^2 r^2}. \label{eq:202}
\end{equation}
The Hamiltonian governing the evolution of the wavepacket is the 
kinetic energy operator $-(1/2\mu)\nabla^2$; we find that
\begin{equation}
C(t)=\left( \frac{2}{2+it/t_0}\right)^{3/2},  \label{eq:203}
\end{equation}
where we recall that $t_0$ characterizes the time scale for the 
evolution of the wavepacket $|\psi \rangle$, and is defined in the 
present example as $t_0=\mu/\kappa_0^2$. Note that $C(t)$ has branch points 
at $t=2it_0$ and $t=\infty$. 
At time $t=0$ the wavepacket has a characteristic width in position space of 
$1/\kappa_0$, but since the wavepacket has a momentum distribution of 
width $\kappa_0$ its spatial spread after time $t$ is 
$\kappa_0 t/\mu$, and this spread exceeds the original width 
of the wavepacket when $t$ is comaparable to $t_0$. In other words, 
the singularity at $2it_0$ is a signature of the time at which 
the wavepacket becomes significantly deformed by spreading.  
The wavepacket can evolve either forwards or backwards in time to the 
{\em single} point $t=\infty$, and the result depends on the arrow of 
time. Hence the wavepacket is double-valued at $t=\infty$, a property 
that is encompassed by the branch point at $t=\infty$. 
 
A general (nonGaussian) wavepacket has a momentum distribution 
whose very-high-momentum components are appreciable, and the spatial 
tail of the wavepacket deforms the moment it begins to 
evolve freely. Even the tail of a Gaussian wavepacket deforms instantly 
if it evolves under the influence of a potential that can transfer 
a large momentum to the particle. Hence, in general $t_0=0$, i.e. 
the correlation amplitude has a branch point singularity at the 
origin. To be more concrete, suppose that $H$ has 
bound-state and scattering-state eigenvectors 
$|\chi_{{\rm bd},n} \rangle$ and $|\chi_{\bf k} \rangle$, 
respectively, with real energy eigenvalues $E_{{\rm bd},n}$ and 
$E_k\equiv k^2/2\mu$, respectively. We can express $|\psi\rangle$ 
as the superposition 
\begin{equation}
|\psi \rangle = \sum_n {\psi}_{{\rm bd},n}
|\chi_{{\rm bd},n} \rangle +
\int d^3 k \; { \psi}({\bf k})|\chi_{\bf k} \rangle,  \label{eq:204}
\end{equation}
where the eigenvectors are normalized so that 
${\psi}_{{\rm bd},n}=\langle \chi_{{\rm bd},n}|\psi\rangle$ 
and ${ \psi}({\bf k})=\langle \chi_{\bf k}|\psi\rangle$. 
This superposition evolves in time as
\begin{equation}
|\psi (t)\rangle = \sum_n{ \psi}_{{\rm bd},n}
|\chi_{{\rm bd},n} \rangle e^{-iE_{{\rm bd},n}t}+
\int d^3 k\; {\psi}({\bf k})e^{-iE_kt}|\chi_{\bf k} \rangle,  \label{eq:205}
\end{equation}
and the correlation amplitude is
\begin{equation}
C(t)= C_{\rm bd}(t)+C_{\rm cont}(t), \label{eq:206}
\end{equation}
where
\begin{eqnarray}
C_{\rm bd}(t) &=& \sum_n |{ \psi}_{{\rm bd},n}|^2 e^{-iE_{{\rm bd},n}t}, 
\label{eq:207}\\
C_{\rm cont}(t) &=& \int d^3 k\; |{\psi}({\bf k})|^2e^{-iE_kt}. \label{eq:208}
\end{eqnarray}
We can let $t$ move into the lower-half complex $t$-plane; the 
exponential $e^{-iE_kt}$ decays with increasing $k$ and both 
$C_{\rm bd}(t)$ and $C_{\rm cont}(t)$ are well-defined. However, 
$e^{-iE_{{\rm bd},n}t}$ explodes as $t\rightarrow \infty$ in the 
lower-half of the $t$-plane; hence $C_{\rm bd}(t)$ has an essential 
singularity at $t=\infty$ and is unbounded. 
Furthermore, if we allow $t$ to move into the 
upper-half complex $t$-plane, $e^{-iE_kt}$ explodes with increasing $k$, 
as the Gaussian $\exp(E_k\; \mbox{Im}\; t)$. Therefore, unless 
$|{ \psi}({\bf k})|^2$ decreases more rapidly than this, $C_{\rm cont}(t)$ 
is formally undefined in the upper-half of the $t$-plane.  
Suppose for the moment that  $|{ \psi}({\bf k})|$ were to decrease 
with increasing $k$ as $e^{-E_kt_0}$. In this case $C_{\rm cont}(t)$ 
would be formally defined in the region $\mbox{Im}\; t<2t_0$. However, 
by rotating the contour of $k$-integration through an angle $\Theta$ 
into the first octant of the lower-right quadrant of the $k$-plane, 
assuming that $|{ \psi}({\bf k})|$ is free of singularities in this 
octant, we can analytically continue $C_{\rm cont}(t)$ throughout 
any finite region of the sector $0\leq \mbox{arg}(t)<2\Theta$ in the 
upper-right quadrant of the $t$-plane, {\em excluding} 
the section of the positive imaginary $t$-axis above $2it_0$ 
since both $e^{-E_kt_0}$ and $e^{-iE_kt}$ are undamped oscillatory 
functions of $k$ when $\Theta=\pi/4$ and $t$ is pure imaginary. 
Therefore $C_{\rm cont}(t)$, and hence $C(t)$, are analytic in both 
the lower- and right-half $t$-planes, but they have 
branch points at $2it_0$, and branch cuts extending from $2it_0$ 
to infinity. In general, $|{ \psi}({\bf k})|$ decreases as a power 
of $1/k$ with increasing $k$, less rapidly than a Gaussian, and 
we may expect $C(t)$ to have a branch point at the origin, i.e. a   
branch point at $2it_0$ with $t_0\rightarrow 0$. Furthermore, 
$|{ \psi}({\bf k})|$ generally has singularities at finite points 
in the complex $k$-plane (in contrast, $e^{-E_kt_0}$ has an 
essential singularity at $k=\infty$); but if there were no 
singularities in the lower-right quadrant of the $k$-plane, 
we could rotate the contour of $k$-integration so that it runs 
along the right edge of the negative imaginary axis, and we 
could subsequently move $t$ from the positive real axis through 
the upper-half $t$-plane to the upper edge of the negative real axis. 
In this case $C(t)$ would be defined everwhere in the finite complex 
$t$-plane, except possibly for a branch point at the origin, 
which we cannot exclude unless we can move $t$ continuously along 
a closed loop around the origin (without discontinuously moving the 
contour of $k$-integration). Hereafter we formally define $t_0$ such that 
{\em the singularity of} $C(t)$ {\em nearest to the origin is located 
at} $2it_0$, and we draw a cut along the positive imaginary axis from 
$2it_0$ to $\infty$. Of course, if, by this definition, $t_0$ were 
to vanish, it could not serve as a unit of time; however, this is not 
of concern in practice, for in practice $t_0$ 
may be small but it does not vanish, a matter we return to at the 
end of this section. 

To determine the general asymptotic behavior of $C(t)$ for $t\sim \infty$ 
let us transform the integration variable ${\bf k}$ to ${\bf k}/\sqrt{t}$ 
on the right side of Eq.~(\ref{eq:208}). Assuming the potential is 
short-range, we can, as a first approximation, replace 
$|{\psi}({\bf k}/\sqrt{t})|^2$ by $|{\psi}({\bf 0})|^2$ for 
$t\sim \infty$. However, if the wavepacket carries orbital angular 
momentum $|{\psi}({\bf 0})|^2$ may vanish; small values of 
$k$ are inhibited by the centrifugal barrier. If $l$ is the smallest 
angular momentum quantum number present in the wavepacket, 
$|{\psi}({\bf k})|^2$ vanishes as $k^{2l}$ for $k\sim 0$. Therefore we write 
\begin{equation}
|{\psi}({\bf k})|^2 \equiv k^{2l}|\psi_l({\bf k})|^2. \label{eq:209b}
\end{equation}
It follows that 
\begin{equation}
C_{\rm cont}(t)\sim e^{-i(2l+3)\pi/4}(2\pi)^{3/2}(2l+1)!! 
|{\psi}_l({\bf 0})|^2 (\mu /t)^{2l+3/2}, \;\;\; t\sim \infty , \label{eq:209}
\end{equation}
where we used 
\[
\int_0^{\infty}dk\; k^{2m}e^{-ak^2}=\frac{(2m-1)!!}{2(2a)^m}
\sqrt{\frac{\pi}{a}}. 
\]
Note that the large-$t$ behavior of $C_{\rm cont}(t)$ is determined 
by small $k$, i.e. by the continuum eigenvalues of $H$ that are 
close to threshold. 
If the potential has a Coulomb tail $|{\psi}({\bf 0})|^2$ is 
infinite, and we must factor out the divergence arising from the 
normalization constant. Thus, if the Coulomb tail is $-Ze^2/r$ at large 
radial distances $r$, we remove an offending factor from 
$|{\psi}({\bf k})|^2$ by writing (for all $l$) 
\begin{equation}
|{\psi}({\bf k})|^2 \equiv \left(\frac{2\pi \gamma}{1-e^{-2\pi \gamma}}
\right) |{\tilde \psi}({\bf k})|^2, \label{eq:210}
\end{equation}
where $\gamma =Z/(a_0k)$, with $a_0=1/\mu e^2$. After making the 
transformation ${\bf k}\rightarrow {\bf k}/\sqrt{t}$, we have 
$\gamma \sim (Z/a_0k)\sqrt{t}$ and the prefactor 
on the right side of Eq.~(\ref{eq:210}) 
either vanishes exponentially (if $Z<0$) or becomes the divergent 
function $(2\pi Z/a_0k)\sqrt{t}$ (if $Z>0$) 
when we let $t$ increase to infinity in any sector excluding the 
negative real axis. Hence, when an attractive ($Z>0$) Coulomb tail is 
present, $C_{\rm cont}(t)$ falls off only as $1/t$: 
\begin{equation}
C_{\rm cont}(t)\sim -8i\pi^2 Z\mu|{\tilde \psi}({\bf 0})|^2/(a_0t), 
\;\;\; t\sim \infty \;\;\; (Z>0). \label{eq:211}
\end{equation}
Note, however, that a Coulomb tail gives rise to an essential singularity 
at $t=\infty$. (If we were to let $t$ increase to infinity in any sector 
excluding the negative real axis on the {\em second} sheet of the 
Riemann $t$-surface, $C_{\rm cont}(t)$ would vanish exponentially 
if $Z>0$.) A branch point at $t=\infty$ remains, but it does not 
dominate the asymptotic behavior. Futhermore, if $Z>0$, Rydberg states 
converging to threshold from below cannot be distinguished in the 
wavepacket from continuum eigenstates converging to threshold 
from above, and therefore we have $C_{\rm bd}(t)\sim C_{\rm cont}(t)$ for 
$t$ approaching $\infty$ in the upper-half $t$-plane (the half-plane 
in which lower-lying bound states decay exponentially). It follows that, 
in the upper-half $t$-plane,
\begin{equation}
C(t)\sim -16i\pi^2 Z\mu|{\tilde \psi}({\bf 0})|^2/(a_0t), 
\;\;\; t\sim \infty \;\;\; (Z>0) . \label{eq:211b}
\end{equation}
In other words, we gain a factor of 2 in the asymptotic form 
of $C(t)$ through the contribution of high Rydberg states. As 
seen in Eq.~(\ref{eq:449d}) of Appendix B, this factor of 2 enters 
the rate for a continuum process near threshold. When the 
potential has a Coulomb tail, the sum (over 
bound states) on the right side of Eq.~(\ref{eq:207}) is convergent, 
but not uniformly, for $t$ in the upper-half $t$-plane, and 
we cannot interchange the limit $t\rightarrow \infty$ and the sum. 
In contrast, when the potential is short-range $C_{\rm bd}(t)\sim 0$ 
for $t\sim \infty$ in the upper-half $t$-plane.


\begin{figure}
  \begin{center}
 \centerline{\psfig{file=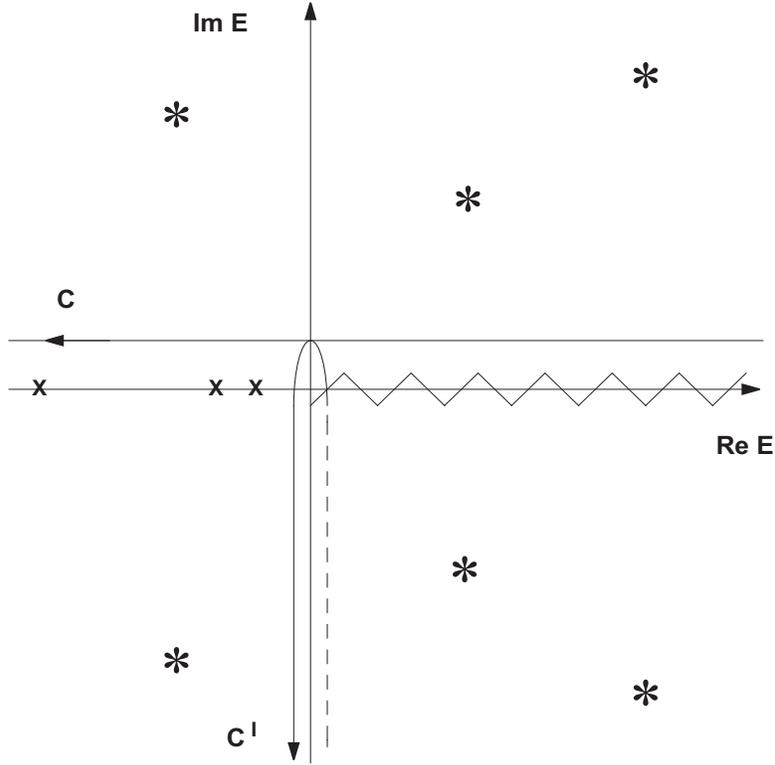,width=4in,height=4in}}
  \end{center}
\caption{The contour ${\cal C}$ runs along the upper edge of the real 
axis of the physical sheet of the Riemann energy surface 
cut (the zig-zag line) along the positive real axis (the ``unitarity'' 
cut). Bound state poles, indicated by $\times$, lie on the negative real axis 
of the physical sheet, while resonance poles, indicated by $\star$, 
lie on the unphysical sheet reached by crossing the cut. The resonance 
poles are distributed 
symmetrically about the cut. The contour ${\cal C'}$ is the result 
of bending ${\cal C}$ around the negative imaginary axis; the 
left half of ${\cal C'}$ lies on the physical sheet, while the 
right half lies on the unphysical sheet. \label{fig1}}
\end{figure}


To gain further insight into the behavior of $C(t)$ at large $t$ it 
is useful to express the time-evolution operator $U(t)$ in terms of 
the resolvent $G(E)$. We have \cite{Gold}
\begin{equation}
U(t)=\frac{1}{2\pi i}\int _{\cal C} dE \; e^{-iEt}G(E), \label{eq:249}
\end{equation}
where, assuming that $t$ is real {\em and} positive, the contour 
${\cal C}$ runs along the upper edge of the real $E$-axis from 
$\infty$ to $-\infty$; see Fig.\ 1. The resolvent has 
branch points at $E=0$ and $E=\infty$, and is defined 
on a two-sheeted Riemann energy surface; the ``unitarity'' 
branch cut is drawn along the positive real $E$-axis.  
In addition to branch points, $G(E)$ has bound-state poles at 
points $E_{{\rm bd},n}$ on the negative real energy axis of one 
sheet --- the physical sheet --- and has resonance poles 
at points $E_{{\rm res},n}$  and $E_{{\rm res},n}^*$ in the lower and 
upper half-planes, respectively, of the 
other sheet --- the unphysical sheet. Let us bend the contour 
${\cal C}$ around the branch point at $E=0$, into the lower-half 
$E$-plane, so that the new contour, ${\cal C'}$, 
wraps around the negative imaginary axis. \cite{Gold} As we 
distort ${\cal C}$ it sweeps over the bound-state poles (on the physical 
sheet) and also over those resonance poles dispersed in the lower-right 
quadrant on the unphysical sheet of the energy-plane. It follows that 
\begin{eqnarray}
|\psi (t)\rangle &=& \frac{1}{2\pi i}\int _{\cal C} dE \; e^{-iEt}G(E)
|\psi \rangle  \label{eq:250}  \\
&=& \sum_n{\psi}_{{\rm bd},n}
|\chi_{{\rm bd},n} \rangle e^{-iE_{{\rm bd},n}t}+
\sum_n{ \psi}_{{\rm res},n}|\chi_{{\rm res},n} \rangle 
e^{-iE_{{\rm res},n}t}+| \psi_{\rm bg}(t)\rangle , \label{eq:251}
\end{eqnarray}
where $|\chi_{{\rm res},n} \rangle$ is an eigenvector of $H$ 
satisfying outgoing-wave boundary conditions corresponding to 
a compound (resonance) state\footnote{
When $E_k$ is complex the continuum eigenfunctions 
$\langle {\bf x} |\chi_{\bf k} \rangle$ explode exponentially in 
position space, as $\exp(i{\bf k}\cdot {\bf x})$, for 
$|{\bf x}|\sim \infty$. Consequently, certain integrals over 
${\bf x}$ --- for example, $\langle \psi |\chi_{{\rm res},n} \rangle$ 
--- are formally undefined. However, such integrals may be defined through 
analytic continuation, e.g. consider $\int_0^{\infty}dr\; e^{ar}$, 
which is $-1/a$ for all $a\neq 0$.}
with a complex energy 
$E_{{\rm res},n}$ whose real part is positive, where 
$\psi_{{\rm res},n}=\langle \chi_{{\rm res},n}
|\psi \rangle$, and where $|\psi_{\rm bg}(t)\rangle$ describes 
the continuum background:
\begin{equation}
| \psi_{\rm bg}(t)\rangle =
\frac{1}{2\pi i}\int _{\cal C'} dE \; e^{-iEt}G(E)
 |\psi \rangle . \label{eq:252}
\end{equation}
As expected from our earlier discussion, if the potential is 
short-range the vector $|\psi_{\rm bg}(t)\rangle$ attenuates as 
$t^{-3/2}$ with 
increasing $t$; a general proof, due to Zumino, is presented in 
Ref.~\cite{Gold}. In the case where the potential has a Coulomb tail, 
the behavior of $|\psi_{\rm bg}(t)\rangle$ at large $t$ has been 
analyzed by Dollard.\cite{Doll} Note that once ${\cal C}$ has been 
deformed to ${\cal C'}$ we can analytically continue the right side 
of Eq.~(\ref{eq:252}), and hence $C(t)$, from the real positive 
$t$-axis to the entire right half of the complex $t$-plane. 

When $t$ is real {\em and} negative, $U(t)$ has an integral representation 
similar to the right side of Eq.~(\ref{eq:249}) but with a contour 
${\bar {\cal C}}$ running along the lower edge of the real $E$-axis 
and in the direction opposite to ${\cal C}$. 
Assuming that the Hamiltonian is invariant under time-reversal, 
the two representations are related through
\begin{equation}
G(E^*)=KG(E)K^{\dagger},  \label{eq:261}
\end{equation}
where $K$ is the antiunitary time-reversal operator.\cite{Mes}
Hence, if $|{\bar \psi} \rangle$ represents the time-reverse of the 
wavepacket $|\psi \rangle$, i.e. if 
$|{\bar \psi} \rangle=K|\psi \rangle$, we have, with 
$t$ real and negative, 
\begin{eqnarray}
|{\bar \psi} (t)\rangle &=& \frac{1}{2\pi i}\int _{\bar {\cal C}} dE \; 
e^{-iEt}G(E)|{\bar \psi} \rangle  \label{eq:262}  \\
&=& \frac{1}{2\pi i}\int _{\bar {\cal C}} dE \; e^{-iEt}KG(E^*)K^{\dagger}
|{\bar \psi} \rangle  \label{eq:263}  \\
&=& K\frac{1}{2\pi i}\int _{\cal C} dE \; e^{iEt}G(E^*)
| \psi \rangle  \label{eq:264}  \\
&=& K| \psi (-t)\rangle,  \label{eq:265}  
\end{eqnarray}
where in the second step we used Eq.~(\ref{eq:261}) and in the 
third step we noted that $K^{\dagger}K=1$ and that $K$ 
complex-conjugates c-numbers. If on the right side of Eq.~(\ref{eq:262}) 
we deform the integration contour into one wrapped around the 
positive imaginary $E$-axis, we obtain an expression for 
$|{\bar \psi}(t)\rangle$ that is similar to the right-side of 
Eq.~(\ref{eq:251}) but with all terms time-reversed. The 
time-reversed resonance terms correspond to the conjugate poles of 
$G(E)$ in the upper-right quadrant of the $E$-plane on the unphysical 
sheet. Let us introduce the new correlation amplitude
\begin{equation}
{\bar C}(t)\equiv \langle {\bar \psi} |{\bar \psi}( t) \rangle. \label{eq:266}
\end{equation}
As long as $t$ is real and negative we can use Eq.~(\ref{eq:265}) 
to write
\begin{eqnarray}
{\bar C}(t) &=& \langle {\bar \psi} |\; [K|\psi (-t) \rangle] \label{eq:267} \\
&=& \langle \psi (-t)|\; [K^{\dagger}|{\bar \psi} \rangle] \label{eq:268} \\
&=& \langle \psi (-t)|\; | \psi \rangle \label{eq:269} \\
&=& [C(-t)]^*,  \label{eq:270}  
\end{eqnarray}
where in the second step we noted \cite{Mes} that, since $K$ is antilinear, 
$\langle b|\; (K|a \rangle)=\langle a|\; (K^{\dagger}|b \rangle)$ 
for any two kets $|a \rangle$ and $|b \rangle$. 
After deforming the contour ${\bar {\cal C}}$, we can analytically continue 
${\bar C}(t)$ into the entire left half of the complex $t$-plane, and since  
both $[{\bar C}(t^*)]^*$ and $C(-t)$ are also analytic functions of $t$ 
in this region we can generalize Eq.~(\ref{eq:270}) to
\begin{equation}
[{\bar C}(t^*)]^*=C(-t) \label{eq:271}
\end{equation}
for $t$ anywhere in the left half of the complex $t$-plane. When 
$t$ lies on the negative imaginary axis, $C(t)$ 
is real, and therefore $[C(-t^*)]^*=C(t)=[{\bar C}(-t^*)]^*$. It 
follows that ${\bar C}(t)$ is the analytic continuation of $C(t)$ 
into the left half of the complex $t$-plane. If the point $t$ is moved 
over the branch cut along the positive imaginary axis, from the left edge of the 
cut to the right edge, on the same sheet, the correlation amplitude jumps 
discontinuously from ${\bar C}(t)$ to $C(t)$; in other words, at the cut 
${\bar C}(t)$ and $C(t)$ are different branches of the same 
multivalued correlation amplitude.

To conclude: In general, $C(t)$ is a nonsingular function of $t$ in the 
{\em right}-half of the finite complex $t$-plane, but has a 
branch point on the positive imaginary axis at $t=2it_0$. We can expand 
$C(t)$ in the right-half of the $t$-plane as --- c.f. Eq.~(\ref{eq:206}) ---
\begin{eqnarray}
C(t) &=& C_{\rm bd}(t)+C_{\rm res}(t)+C_{\rm bg}(t) \label{eq:272} \\
&\equiv & \sum_n |{ \psi}_{{\rm bd},n}|^2 e^{-iE_{{\rm bd},n}t}+
\sum_n |\psi_{{\rm res},n}|^2 e^{-iE_{{\rm res},n}t} +
\langle \psi |\psi_{\rm bg}(t)\rangle, \label{eq:273} 
\end{eqnarray}
where $C_{\rm bg}(t)$ falls off as $t^{-3/2}$ or, if the potential has an 
attractive Coulomb tail, as $1/t$ for $t\sim \infty$. If the potential 
has a Coulomb tail $C_{\rm bg}(t)$ has an essential singularity at 
$t=\infty$. Due to the exponential factor(s), $C_{\rm bd}(t)$ and 
$C_{\rm res}(t)$ also have essential singularities at $t=\infty$. In  
fact, as $t$ increases to infinity in the lower half of the the $t$-plane 
$C_{\rm bd}(t)$ explodes exponentially, and $C_{\rm res}(t)$ exhibits 
similar behavior for $t$ within some other sector of the $t$-plane. 
However, $C_{\rm bd}(t)$ is bounded in the upper-half of the 
$t$-plane. Also, $C_{\rm res}(t)$ is bounded in the 
lower-right quadrant of the $t$-plane since Re~$E_{{\rm res},n}>0$ and 
Im~$E_{{\rm res},n}<0$. In fact, $C_{\rm res}(t)$ is 
bounded in an even wider region since $G(E)$ has no poles within 
a sector of the $E$-plane, say 
$-\Xi_0 <\mbox{arg} \; (E) < \Xi_0$, which contains 
the positive real energy axis; therefore 
$\mbox{Im}\; E_{{\rm res},n}t<0$, and $C_{\rm res}(t)$ is bounded, 
throughout the sector $0 \leq  \mbox{arg}\; (t) < \Xi_0$ of the 
$t$-plane. Hence $C(t)$ is bounded throughout the sector 
$0 \leq  \mbox{arg}\; (t) < \Xi_0$. Similarly, ${\bar C}(t)$ is 
nonsingular in the {\em left}-half of the finite complex $t$-plane, 
${\bar C}_{\rm res}(t)$ is bounded throughout the 
lower-left quadrant of the $t$-plane, and beyond, and 
${\bar C}(t)$ is bounded throughout the 
sector $\pi -\Xi_0 \leq  \mbox{arg}\; (t) < \pi$. We can 
generate a power series for $C(t)$ by expanding $U(t)=e^{-iHt}$ in 
powers of $t$; we have
\begin{equation}
|\psi (t)\rangle =\sum_{m=0}^{\infty} \frac{(-iHt)^m}{m!}
 |\psi\rangle, \label{eq:274}
\end{equation}
and therefore introducing the dimensionless variable 
$\tau=t/t_0$, assuming that $t_0\neq 0$, we obtain
\begin{equation}
C(t)=\sum_{m=0}^{\infty} \langle \psi |(-iHt_0)^m|\psi \rangle 
\frac{\tau^m}{m!}. \label{eq:275}
\end{equation}
The radius of convergence of this expansion of $C(t)$ in powers of 
$\tau$ is 2 since the singularity of $C(t)$ that is nearest 
to the origin is, by definition, located at $2it_0$. Note 
that we just assumed $t_0\neq 0$, yet earlier we remarked that 
we can have $t_0= 0$. While in principle it is generally true 
that $t_0=0$, in practice $t_0\neq 0$, 
as we now explain. The time that it takes the 
wavepacket to deform significantly from its initial form is governed 
by the highest speed in the velocity distribution of the wavepacket 
as it evolves in the presence of the potential. Now the highest 
speed in the velocity distribution is effectively determined by the largest 
eigenvalue, $E_{\rm max}$ say, of the smallest matrix $\underline{H}$ 
that can accurately represent $H$ in the expression 
$e^{-iHt}|\psi \rangle$. Hence, in practice the characteristic value 
of $t_0$ is $1/E_{\rm max}$, which may be very small but is nonzero. \\


\begin{figure}
  \begin{center}
 \centerline{\psfig{file=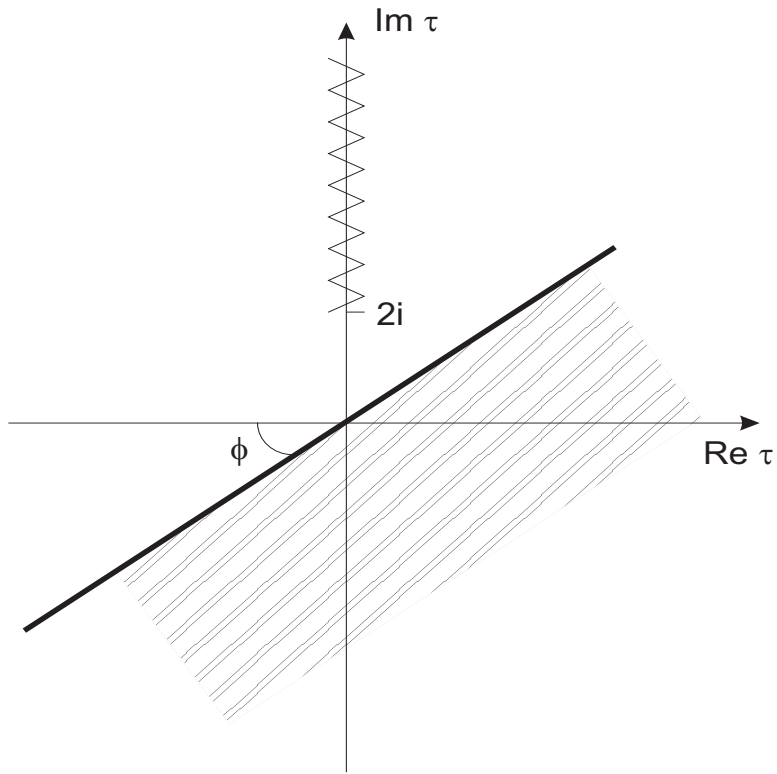,width=4in,height=4in}}
  \end{center}
\caption{The correlation amplitude $C(t)$ has a branch-point singularity 
on the positive imaginary axis in the $\tau \equiv t/t_0$ plane, and 
another one at infinity. The unit of time $t_0$ is defined so that the 
singularity nearest the origin is at $\tau=2i$. We have drawn a cut extending 
up the positive imaginary axis from $2i$. One-half (the hatched section) 
of the $\tau$-plane is conformally mapped onto a unit circle. 
\label{fig2}}
\end{figure}


{\large\bf 3 Conformal Transformation}\\

To calculate the rate for some continuous stationary or quasistationary 
process from a correlation amplitude we need to know the correlation 
amplitude for all $t$ on either the positive or negative real axis. 
However, the power series in $\tau$ is useful only for $0\leq |t| <2t_0$. 
Since $C(t)$ is singular on the positive imaginary axis, 
it is expedient to divide the $t$-plane into two half-planes 
separated by a line through the origin, and to conformally map the 
lower of these half-planes (i.e. the one free of singularities) 
into the unit circle; see Fig.\ 2. Thus we change variables from $t$ to 
\begin{equation}
u=\frac{\tau+ie^{i\phi}}{\tau-ie^{i\phi}}.\label{eq:301}
\end{equation}
The mapping $u$ depends on both $t$ and the complex unit of time
\begin{equation}
t_{\phi}=e^{i\phi}t_0, \label{eq:307}
\end{equation}
and a particular branch of $C(t)$ can be specified by $\phi$, or, 
rather, a range of values of $\phi$. As $\phi$ varies over the range 
$-\pi/2<\phi <\pi/2$ the boundary line 
$\mbox{Im}\; t/\mbox{Re}\; t =\tan \phi$ rotates through one revolution, 
and we remain on the same branch of $C(t)$. To pass to another branch 
of $C(t)$ we must allow $\phi$ to move out of this range. We can 
do this by analytic continuation in the variable $t_{\phi}$, 
to the left-half of the $t_{\phi}$-plane, i.e. to values of 
$\phi$ in the range $\pi/2 \leq |\phi|<\pi$. 

Note that the boundary line $\mbox{Im}\; t/\mbox{Re}\; t =\tan \phi$ 
is mapped onto the circumference of the unit circle in the $u$-plane, 
and the real positive (negative) $t$-axis is mapped onto the broken 
line shown in the upper (lower) semicircle of Fig.\ 3 if $\phi$ is 
positive (negative). The points $t=0$, $t=t_{\phi}$, 
$t=\infty$, $t=-it_{\phi}$, and $t=it_{\phi}$ are mapped 
onto the points $u=-1$, $u=i$, $u=1$, $u=0$, and $u=\infty$, respectively. 
There are no singularities inside the unit circle, and only one 
on the circumference, at $u=1$ (corresponding to the singularity at 
$t=\infty$). A singularity on the positive imaginary axis of the 
$t$-plane is mapped onto a point outside the unit circle in the 
lower (upper) half of the $u$-plane if $\phi$ is positive (negative), 
and moves onto the real $u$-axis as $\phi$ vanishes.


\begin{figure}
  \begin{center}
 \centerline{\psfig{file=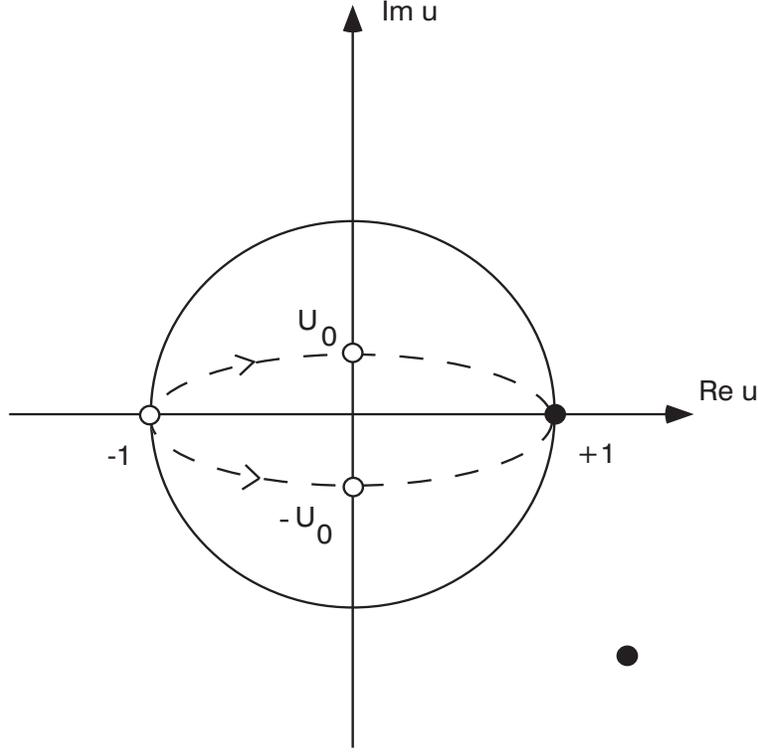,width=4in,height=4in}}
  \end{center}
\caption{The unit circle in the $u$-plane, where 
$u=(\tau+ie^{i\phi})/(\tau-ie^{i\phi})$. If the phase $\phi$   
is positive (negative), the real positive (negative) axis in the 
$\tau$-plane is mapped onto the trajectory indicated by the broken line 
in the upper (lower) semicircle of the $u$-plane. This trajectory is 
furthest from the real $u$-axis at the point $u_0$ or $-u_0$ where $u_0=
i\cos \phi/(1+\sin |\phi|)$. The correlation amplitude has a 
singularity (indicated by $\bullet$) at $u=1$ and another singularity 
outside the unit circle. \label{fig3}}
\end{figure}


We now express $C(t)$ as a power series in $u$. Substituting for 
$\tau$ in the power series on the right side of Eq.~(\ref{eq:275}) using
\begin{equation}
\tau=-ie^{i\phi}\left( \frac{1+u}{1-u}\right), \label{eq:302}
\end{equation}
and noting that
\begin{equation}
\left( \frac{1+u}{1-u}\right)^m
=\sum_{n=0}^{\infty} \frac{1}{n!} P_n(m)u^n,  \label{eq:303}
\end{equation}
where $P_n(m)$ is the polynomial\footnote{
Incidentally, it can be shown that 
\[
P_n(m)=\frac{(m+n-1)!}{(m-1)!}\; _2F_1(-n,-m,-m-n+1;-1),\;\;\; n\geq 1, 
\]
with $P_0(m)=1$. The $P_n(m)$ are Krawtchouk polynomials, which 
for $n\geq 1$ satisfy the recursion relation
\[
P_n(x)=2xP_{n-1}(x)+(n-1)(n-2)P_{n-2}(x),
\]
with $P_0(x)=1$.}
\begin{equation}
P_n(m)=\left[ \frac{d^n}{du^n}\left( 
\frac{1+u}{1-u}\right)^m \right]_{u=0},  \label{eq:304}
\end{equation}
we obtain the new power series:
\begin{equation}
C(t)=\sum_{n=0}^{\infty} c_n(t_{\phi}) u^n, \label{eq:308}
\end{equation}
where
\begin{equation}
c_n(t_{\phi})=\langle \psi |C_n(-t_{\phi}H)|\psi \rangle, \label{eq:309}
\end{equation}
and where, with $z$ the operator $-t_{\phi}H$,
\begin{equation}
C_n(z)=\sum_{m=0}^{\infty} \frac{P_n(m)}{n!m!}z^m.  \label{eq:310}
\end{equation}
Since $u=-1$ when $t=0$, and since $C(0)= \langle \psi |\psi \rangle$, 
we have
\begin{equation}
\sum_{n=0}^{\infty} (-1)^n c_n(t_{\phi})=\langle \psi |\psi \rangle, 
\label{eq:311}
\end{equation}
which provides a useful check on the accuracy of the coefficients 
$c_n(t_{\phi})$ in a practical application. 
Since $C(t)$ is nonsingular everywhere inside the unit 
circle in the $u$-plane, the power series in $u$, on the right side of 
Eq.~(\ref{eq:308}), converges for all $|u|<1$. The point $u=1$ 
requires special consideration, as discussed below.

The series for $C_n(z)$ can be expressd in closed form. This is 
obvious when $n=0$ since $P_0(m)=1$ and we have 
\begin{equation}
C_0(z)=e^z. \label{eq:312}
\end{equation}
For $n\geq 1$, we use Eqs.~(\ref{eq:304}) and (\ref{eq:310}) to write:
\begin{eqnarray}
C_n(z)&=&\frac{1}{n!}\left\{\frac{d^n}{du^n}\left [\sum_{m=0}^{\infty}
\frac{z^m}{m!}\left (\frac{1+u}{1-u}\right )^m\right ]\right\}_{u=0},
\nonumber \\ \nonumber \\
 &=&\frac{1}{n!}e^z\left[\frac{d^n}{du^n}e^{2z\left (\frac{u}{1-u}\right )}
\right ]_{u=0},  \label{eq:314}
\end{eqnarray}
where in the second step we noted 
\[
e^{z\left (\frac{1+u}{1-u}\right )}=e^z e^{2z\left (\frac{u}{1-u}\right )}.
\]
Now the generating function of the 
ordinary Laguerre polynomials, $L_m(x)$, is\cite{Abra}
\begin{equation}
(1-u)^{-1}e^{-x\left (\frac{u}{1-u}\right )}=\sum_{m=0}^{\infty} 
L_m(x)u^m, \label{eq:315}
\end{equation}
and it follows that
\begin{equation}
C_n(z)=e^z[L_n(-2z)-L_{n-1}(-2z)]. \label{eq:316}
\end{equation}
Since $L_n(x)-L_{n-1}(x)=-(x/n)L^{(1)}_{n-1}(x)$, where $L^{(1)}_{n-1}(x)$ 
is an associated Laguerre polynomial of degree $n-1$, which may be recast 
as a confluent hypergeometric function $n_1F_1(1-n,2,x)$, we arrive at 
\begin{equation}
C_n(z)=(2z)\;_1F_1(1-n,2,-2z)e^z,\; \; n\geq 1, \label{eq:317}
\end{equation}
which is the desired result.

It is instructive to write $C_n(z)\equiv Q_n(z)e^z$, where $Q_n(z)$ is 
a polynomial of degree $n$ in $z$, so that 
\begin{equation}
c_n(t_{\phi})=\langle \psi |Q_n(-t_{\phi}H)|\psi (-it_{\phi})\rangle, 
\label{eq:318}
\end{equation}
where $|\psi (-it_{\phi})\rangle$ is the wavepacket that has evolved 
for the complex time $-it_{\phi}$ from the wavepacket $|\psi \rangle$ 
at $t=0$; since the point $t=-it_{\phi}$ is further from the singularity 
at $2it_0$ than the point $t=0$, the influence of the singularity 
on $|\psi (-it_{\phi})\rangle$ is weaker than on $|\psi \rangle$. 
The point $t=-it_{\phi}$, or, equivalently, 
$\tau=-ie^{i\phi}$, corresponds to the 
origin of the $u$-plane, i.e. the point about which a power series 
in $u$ is developed. Hence the conformal 
transformation permits the analytic continuation of $C(t)$ from a 
power series in $\tau$ to a power series in $u$ {\em via} 
a ``connection'' point $\tau=-ie^{i\phi}$ which lies 
within the circle of convergence of the power series 
in $\tau$, but on the side of this circle furthest from the 
singularity at $2it_0$. 

We now explore some properties of the expansion coefficients 
$c_n(t_{\phi})$. It is evident 
from Eqs.~(\ref{eq:309}), (\ref{eq:312}), and (\ref{eq:317}) that 
since $H$ is Hermitian, and since $C_n(z)$ is a real function when 
$z$ is real, 
\begin{equation}
[c_n(t^*_{\phi})]^*=c_n(t_{\phi}).  \label{eq:309b}
\end{equation}
Furthermore, since $t^*_{\phi}=t_{-\phi}$ we have
\begin{equation}
c_n(t_{-\phi})=[c_n(t_{\phi})]^*  . \label{eq:309c}
\end{equation}
We know from the previous 
section that $\langle \psi |e^{-t_{\phi}H}|\psi \rangle$ is analytic 
in the $t_{\phi}$-plane except on the negative imaginary axis, where 
$\phi=\pm \pi$. Hence $c_0(t_{\phi})$ is analytic everywhere 
in the finite $t_{\phi}$-plane cut along the negative imaginary axis. 
Since $Q_n(-t_{\phi}H)$ is a polynomial, and therefore an analytic function 
of $t_{\phi}H$, we infer that $c_n(t_{\phi})$ is also analytic 
everywhere in the finite cut $t_{\phi}$-plane, where $n$ is any 
nonnegative integer. 

Although the expansion coefficients $c_n(t_{\phi})$ are analytic throughout 
the range $|\phi|<\pi$, if we allow $\phi$ to move out of the range 
$|\phi|<\pi/2$ the boundary line $\mbox{Im}\; t/\mbox{Re}\; 
t =\tan \phi$ crosses the cut that we have drawn along the 
positive imaginary $t$-axis. Hence if $\pi/2\leq |\phi| <\pi$
this cut is mapped into the unit circle and the 
expansion of $C(t)$ in powers of $u$, i.e. Eq.~(\ref{eq:308}), 
no longer converges for all $u$ within the unit circle. 
Nevertheless, provided that $u$ is sufficiently small, i.e. 
provided that $t$ is sufficiently close to $-it_{\phi}$, the 
expansion of $C(t)$ in powers of $u$ converges even for 
$\pi/2\leq |\phi| <\pi$. As $\phi$ increases from 0 to $\pi_-$ --- 
where in general $a_-$ and $a_+$, respectively, are numbers just 
below and just above the number $a$ ---  the 
point $-it_{\phi}$ moves from the negative imaginary axis, into the 
right-half of the $t_{\phi}$-plane, and onto right edge of the positive 
imaginary axis. Thus, there is always a region in the right-half of the 
$t$-plane within which the expansion of $C(t)$ in powers of $u$ converges 
when $0\leq \phi <\pi$; but as $\phi$ approaches $\pi_-$ this 
region shrinks to the point $i$ on the cut in the $t$-plane. Similarly, 
as $\phi$ decreases from 0 to $-\pi_-$, the 
point $-it_{\phi}$ moves from the negative imaginary axis, into the 
left-half of the $t_{\phi}$-plane, and onto left edge of the positive 
imaginary axis. Thus, there is always a region in the left-half of the 
$t$-plane within which the expansion of $C(t)$ in powers of $u$ converges 
when $-\pi<\phi \leq 0$; but as $\phi$ approaches $-\pi_-$ this 
region again shrinks to the point $i$ on the cut. Recall that ${\bar C}(t)$ 
is the analytic continuation of $C(t)$ from the right-half to the 
left-half of the $t$-plane. As long as $|\phi|<\pi/2$ we can 
express ${\bar C}(t)$ as a convergent power series in $u$ for all $u$ 
inside the unit circle; in analogy with Eq.~(\ref{eq:308}) we have 
\begin{equation}
{\bar C}(t)=\sum_{n=0}^{\infty} {\bar c}_n(t_{\phi}) u^n, \label{eq:308b}
\end{equation}
where
\begin{equation}
{\bar c}_n(t_{\phi})=\langle {\bar \psi} |C_n(-t_{\phi}H)
|{\bar \psi} \rangle .  \label{eq:309d}
\end{equation}
Recalling that $|{\bar \psi} \rangle =K| \psi \rangle$ and that 
$K$ is the antiunitary time-reversal operator, so that $K^{\dagger}K=1$ and 
$(\langle b|K^{\dagger})|a\rangle=[\langle b|(K^{\dagger}|a\rangle)]^*$, 
we infer that for $|\phi|<\pi/2$ we have 
\begin{equation}
{\bar c}_n(t_{\phi})=c_n(t_{\phi}).  \label{eq:309e}
\end{equation}
This identity is hardly surprising since $C(t)$ and ${\bar C}(t)$ 
represent the same (multivalued) function of $t$, and as long as 
$|\phi|<\pi/2$ there is a {\em common} region in the lower-half of the 
$t$-plane within which the expansions of both $C(t)$ and ${\bar C}(t)$ 
in powers of $u$ converge. However, once $\phi$ crosses the line 
$|\phi|=\pi/2$ into the region $|\phi|\geq \pi/2$ the characters of 
$c_n(t_{\phi})$ and ${\bar c}_n(t_{\phi})$ differ since 
there is no common region in the $t$-plane within which 
the expansions of $C(t)$ and ${\bar C}(t)$ both converge. On the 
cut along the negative imaginary axis in the $t_{\phi}$-plane, 
$c_n(t_{\pi})$ and ${\bar c}_n(t_{-\pi})$ are different branches of  
the same multivalued function of $t_{\phi}$.

An alternative form for $C_n(z)$, from which we may deduce 
the behavior of the coefficients $c_n(t_{\phi})$ for large $n$, can 
be obtained using a standard expansion of the confluent hypergeometric 
function in terms of Bessel functions.\cite{Abra} We find that 
\begin{equation}
C_n(z)=-2\sum_{m=0}^{\infty}\; A_m(n)B_{m+1}(z),
\; \; n\geq 1, \label{eq:319}
\end{equation}
where 
\begin{equation}
B_m(z)=(-z/2n)^{m/2}J_m(\sqrt{-8nz}), \label{eq:320}
\end{equation}
with $J_n(z)$ the regular Bessel function, and where 
$A_0(n)=1$, $A_1(n)=0$, $A_2(n)=1$, and 
\begin{equation}
A_{m+1}(n)=A_{m-1}(n)-[2n/(m+1)]A_{m-2}(n),\; \; \; m\geq 2. \label{eq:321}
\end{equation}
For large $n$ and for $z$ fixed, real and negative (so that 
$\sqrt{-z}$ is real and positive) we have, using the asymptotic form 
of $J_m(x)$ for large $x$, 
\begin{equation}
B_m(z)\sim \left( \frac{-z^{2m-1}}{2^{2m+1}\pi^2 n^{2m+1}}\right)^{1/4}
\cos(\sqrt{-8nz}-\frac{1}{2}m\pi-\frac{1}{4}\pi), \;\;\; 
n \sim \infty.  \label{eq:331}
\end{equation}
It follows that for $n \sim \infty$, and $z$ fixed, real and negative, 
the first term in the series on the right side of Eq.~(\ref{eq:319}) 
dominates, and we have
\begin{equation}
C_n(z)\sim \left( \frac{-2z}{\pi^2 n^3}\right)^{1/4}
\cos(\sqrt{-8nz}+\pi/4), \;\;\; n \sim \infty. \label{eq:339}
\end{equation}
Of course, $-z$ is not a fixed number, but rather is the operator 
$t_{\phi}H$, which has a spectrum consisting of a discrete set of 
negative eigenvalues and a continuum of positive eigenvalues. 
From Eqs.~(\ref{eq:204}) and (\ref{eq:309}) we have
\begin{equation}
c_n(t_{\phi})= c_{{\rm bd},n}(t_{\phi})+c_{{\rm cont},n}(t_{\phi}),
\label{eq:342}
\end{equation}
where, using Eq.~(\ref{eq:339}), 
\begin{eqnarray}
c_{{\rm bd},n}(t_{\phi}) &\equiv &\sum_m |\psi_{{\rm bd},m}|^2 
C_n(-t_{\phi}E_{{\rm bd},m}) \label{eq:348} \\
&& \sim \left(\frac{2}{\pi^2 n^3}\right)^{1/4}
 \sum_m |\psi_{{\rm bd},m}|^2 (t_{\phi}E_{{\rm bd},m})^{1/4}
\cos(\sqrt{8nt_{\phi}E_{{\rm bd},m}}+\pi/4),  \label{eq:343} 
\end{eqnarray}
and
\begin{eqnarray}
c_{{\rm cont},n}(t_{\phi}) &\equiv & \int d^3 k\; 
|{\psi}({\bf k})|^2C_n(-t_{\phi}E_k)  \label{eq:344} \\
&& \sim  \left(\frac{2}{\pi^2 n^3}\right)^{1/4}
\int d^3 k\; (t_{\phi}E_k)^{1/4}|{\psi}({\bf k})|^2
\cos(\sqrt{8nt_{\phi}E_k}+\pi/4)e^{-\eta k},  \label{eq:345} 
\end{eqnarray}
where in the last step we inserted an unobtrusive factor of $e^{-\eta k}$ 
(with $\eta$ positive but infinitesimal) to ensure convergence 
at a later stage. We choose the phase of $E_{{\rm bd},m}$ to be $\pi$, 
rather than $-\pi$, i.e. we write $E_{{\rm bd},m}=e^{i\pi}|E_{{\rm bd},m}|$, 
since the bound state poles are reached from the upper ``physical'' edge 
of the unitarity cut by following a path in the upper-half of 
the energy plane. Note that while $J_m(\sqrt{-8nz})$ has a branch point 
singularity at $z=0$, due to the square root in the argument, $B_m(z)$ 
does not have a branch point at $z=0$ since $J_m(x)$ is 
proportional to $x^m$ for $x\sim 0$. On the other hand, 
if we take the liberty of using the asymptotic form of $B_m(z)$ --- see 
Eq.~(\ref{eq:331}) --- for complex values of $z$, we infer 
that $B_m(z)$ is not single-valued when 
$\phi$ is varied, with $|z|$ held fixed, around a closed loop from 
$-\pi$ to $\pi$, and indeed we know that $c_n(t_{\phi})$ is not 
single-valued when $\phi$ is varied from $-\pi$ to $\pi$.

Since $E_{{\rm bd},m}<0$ the bound-state terms 
on the right side of Eq.~(\ref{eq:343}) explode exponentially for 
$n\sim \infty$, unless $\phi=\pm \pi$. Hence $c_{{\rm bd},n}(t_{\phi})$ 
also explodes exponentially, unless $\phi=\pm \pi$ (excluding 
a potential with a Coulomb tail, a case considered below.) 
This singular behavior is related to the fact that 
$C_{\rm bd}(t)$ has an essential singularity at $t=\infty$, 
i.e. at $u=1$. As $t$ approaches $\infty$ in the lower-half $t$-plane, 
$C_{\rm bd}(t)$ explodes exponentially, and unless $\phi=\pm \pi$ 
there is a nonvanishing sector of the lower-half $t$-plane within 
which $t$ approaches $\infty$ and, concomitantly, $u$ approaches 
unity {\em within} the unit circle (this sector is 
the full half-plane if $|\phi|=0_+$). Therefore, 
unless $\phi=\pm \pi$, the power series in $u$ 
converges at $u=1$ only if we exclude from $|\psi \rangle$ 
the bound-state eigenvectors of $H$. Thus we replace 
$|\psi \rangle$ by $P_{\rm cont}|\psi \rangle$ where 
$P_{\rm cont}=1-P_{\rm bd}$ and where $P_{\rm cont}$ and 
$P_{\rm bd}$ are projection operators with $P_{\rm bd}$ defined as
\begin{equation}
P_{\rm bd}=\sum_m\; |\chi_{{\rm bd},m} \rangle \; \langle \chi_{{\rm bd},m}|, 
\label{eq:340}
\end{equation}
where, if the potential is short-range (i.e. no Coulomb tail), the 
sum is over all bound states. The omission of these bound states does 
not affect the rate for a continuum process (if the potential is 
short-range), but it may affect the energy shift of the system. 
Note that while the compound states also give rise to   
an essential singularity, {\em they need not be omitted}; the sector of the 
$t$-plane in which they explode corresponds to letting 
$u$ approach unity from {\em without} the unit circle, provided 
that $|\phi| < \Xi_0$. 

To obtain the large-$n$ behavior of 
$c_{{\rm cont},n}(t_{\phi})$, we change variables from ${\bf k}$ to 
${\bf k}/\sqrt{n}$ on the right side of Eq.~(\ref{eq:345}).  
Let us first treat the case of a short-range potential, and for 
simplicity let us assume that the angular momentum quantum number $l$ 
is zero. We can factor $|{\psi}({\bf k}/\sqrt{n})|^2$ out of the integral 
as $|{\psi}({\bf 0})|^2$, and we find that
\begin{equation}
c_{{\rm cont},n}(t_{\phi}) \sim \frac{15\pi}{2^{9/2}}
\left(\frac{\mu}{t_{\phi}}\right)^{3/2}
\frac{|{\psi}({\bf 0})|^2}{n^{5/2}},  \;\;\; n\sim \infty  \label{eq:346} 
\end{equation}
where we used
\begin{equation}
\int d^3 k\; (t_{\phi}E_k)^{1/4}\cos(\sqrt{8t_{\phi}E_k}+\pi/4)e^{-\eta k}
=\frac{15\pi^{3/2}}{2^{19/4}}
\left(\frac{\mu}{t_{\phi}}\right)^{3/2}.  \label{eq:347} 
\end{equation}
We pause for a few remarks. First, while the integral on the left side 
of Eq.~(\ref{eq:347}) is formally defined 
only for $t_{\phi}$ real and positive, both sides of Eq.~(\ref{eq:346}) 
are analytic everywhere in the finite $t_{\phi}$-plane cut along the 
negative real axis. Second, if $l\neq 0$ we must modify 
Eq.~(\ref{eq:346}) by including a factor proportional to $1/n^{2l}$. 
Finally, note that the large-$n$ behavior of $c_{{\rm cont},n}(t_{\phi})$ 
is determined by small $k$, i.e. by the continuum eigenvalues of 
$H$ close to threshold. Since the large-$t$ behavior of 
$C_{\rm cont}(t)$ is also determined by the continuum eigenvalues 
close to threshold, we infer that the large-$t$ behavior of 
$C_{\rm cont}(t)$ is determined by those terms with large $n$ in
the power series on the right side of Eq.~(\ref{eq:308}). 

It follows that if the potential is short-range, 
$c_{{\rm cont},n}(t_{\phi})$ decreases as $n^{-(2l+5)/2}$ 
as $n$ increases. We now consider the case where the potential 
has an attractive Coulomb tail. We must first reconsider 
$c_{{\rm bd},n}(t_{\phi})$. There are an infinite number of Rydberg 
bound states converging to threshold for which $|8nt_{\phi}E_{{\rm bd},m}|<1$ 
no matter how large is $n$. Such states do not yield divergent 
terms on the right side of Eq.~(\ref{eq:343}) 
and, as noted earlier, cannot be distinguished from 
continuum states converging to threshold; therefore, they 
should not be projected out of the wavepacket when $\phi \neq \pm \pi$. 
We write $E_{{\rm bd},m}=-Z^2e^2/[2(m^*)^2a_0]$, 
where $m^*=m-\delta$, with $\delta$ the quantum defect, 
which is roughly independent of $m$ for $m\gg 1$. We must 
retain at least those bound states for which $m^*$ is roughly greater 
than or of the order of $\sqrt{n}$. If $m*\gg 1$ we can relate 
$|\psi_{{\rm bd},m}|^2$ to $|{\psi}({\bf k})|^2$ as follows. 
The wavepacket $|\psi\rangle$ contains Rydberg bound states
with population $|\psi_{{\rm bd},m}|^2dm^*$ in the interval 
$(m^*,m^*+dm^*)$, and continuum states with population 
$|{\psi}({\bf k})|^2d^3k$ in the interval $d^3k$ centered at 
${\bf k}$. Just above threshold, where $ka_0\ll Z$, we can replace 
$|{\psi}({\bf k})|^2$ by  $2\pi\gamma |{\tilde \psi}({\bf 0})|^2$, where 
we used  Eq.~(\ref{eq:210}) noting that $e^{-2\pi \gamma}\ll 1$. 
Since $dE_{{\rm bd},m}/dm^*=Z^2e^2/[(m^*)^3a_0]$ and 
$dE_k/dk=k/\mu$, and since the differential populations with respect 
to energy must be the same for bound states just below, and continuum 
states just above, threshold, we have, writing  $d^3k=4\pi k^2dk$,
\begin{equation}
|\psi_{{\rm bd},m}|^2=\left(\frac{8\pi^2 Z^3}{(m^*a_0)^3}\right)
|{\tilde \psi}({\bf 0})|^2.
\label{eq:349d} 
\end{equation}
It follows from Eqs.~(\ref{eq:343}) and (\ref{eq:349d}) that, 
with $t_{\phi}$ on the upper edge of the cut along the negative real 
axis of the $t_{\phi}$-plane, and with $E_{{\rm bd},m}=e^{i\pi}
|E_{{\rm bd},m}|$, 
\begin{equation}
c_{{\rm bd},n}(t_{\phi}) \sim 8\pi^2 Z^3 |{\tilde \psi}({\bf 0})|^2 
\left(\frac{2}{\pi^2 n^3}\right)^{1/4}
 \sum_{m=1}^{\infty} \frac{(t_{\phi}E_{{\rm bd},m})^{1/4}}{(m^*a_0)^3}
\cos(\sqrt{8nt_{\phi}E_{{\rm bd},m}}+\pi/4), 
\label{eq:349e} 
\end{equation}
where we have extended the lower limit of the sum to 1 since 
the leading $n$-behavior of the sum is relatively insensitive to 
the lower limit. In Appendix D we analyse the large-$n$ behavior 
$c_{{\rm bd},n}(t_{\phi})$ and $c_{{\rm cont},n}(t_{\phi})$
for all complex $t_{\phi}$. To summarise the results of 
Appendix D, when the potential has an attractive Coulomb tail we find that 
\begin{equation}
c_n(t_{\phi})  \sim
\frac{i}{n^2}\left( \frac{3\pi^2Z\mu |{\tilde \psi}({\bf 0})|^2}
{2^{5/2}a_0t_{\phi}}\right),  \; \; \; -\pi < \phi <\pi  ,
\label{eq:350c} 
\end{equation}
for $t_{\phi}$ on the first sheet, and  
\begin{equation}
c_n(t_{\phi})  \sim
-\frac{i}{n^2}\left( \frac{3\pi^2Z\mu |{\tilde \psi}({\bf 0})|^2}
{2^{5/2}a_0t_{\phi}}\right),  \; \; \; \pi < \phi <3\pi  ,
\label{eq:350d} 
\end{equation}
for $t_{\phi}$ on the second sheet. Hence, $c_n(t_{\phi})$ 
has not only a branch cut on the negative real $t_{\phi}$-axis, 
but also a discontinuity which amounts to a sign reversal.\\

{\large\bf 4 Green function}\\

The inclusive rate at which a continuous stationary or quasistationary 
process occurs if ${\cal E}$ is the positive energy of the system is 
$-2\mbox{Im}\; R({\cal E}+i\eta)$ where $\eta$ is positive but infinitesimal 
and where $R(E)$ is a Green function matrix element of the form 
\begin{equation}
R(E) \equiv \langle \psi |G(E)|\psi \rangle. \label{eq:400}
\end{equation}
The real part of $R({\cal E}+i\eta)$ includes the energy shift of the system.  
For example, if the system consists of an infinitely heavy particle 
(at rest) and a light particle that is incident from infinity 
in the unperturbed state represented by $|\Psi_0 \rangle$, we obtain 
the rate for the light particle to scatter by putting 
$|\psi \rangle=W|\Psi_0 \rangle$ where $W$ is the interparticle 
potential. As another example, consider a system that consists 
of an atom, initially bound in 
the unperturbed state represented by $|\Psi_0 \rangle$; if this 
atom is exposed to weak monochromatic radiation, we obtain the rate 
for the atom to decay by putting $|\psi \rangle=V_{+}|\Psi_0 \rangle$ 
where $V_{+}$ is the one-photon absorption operator.

Provided that $E$ lies in the upper half of the complex $E$-plane
we can represent $R(E)$ by the integral 
\begin{eqnarray}
R(E) &=& -i\int_0^{\infty}dt\; e^{iEt}C(t) \label{eq:401}\\
&=& -it_0\int_0^{\infty}d\tau \; e^{i(t_0E)\tau}C(t). \label{eq:402}
\end{eqnarray}
Since $C(t)$ is bounded throughout the sector 
$0 \leq  \mbox{arg}\; (\tau) < \Xi_0$ we can rotate the contour of 
integration into this sector and extend 
the integral representation to a sector in the lower-right 
quadrant of the $E$-plane. To analytically 
continue $R(E)$ into a sector of the $E$-plane that includes the 
negative real $E$-axis we first recall that both 
$C_{\rm res}(t)$ and $C_{\rm bg}(t)$ are bounded throughout 
the lower right quadrant of the $\tau$-plane, but $C_{\rm bd}(t)$ 
is unbounded in this quadrant. Therefore, provided that we 
project out the bound states, and choose $E$ to lie temporarily 
on the left edge of the positive imaginary $E$-axis, 
we can rotate the integration contour into the lower-right 
quadrant of the $\tau$-plane so that it runs along the right edge 
of the negative imaginary $\tau$-axis. Subsequently, we can move $E$ 
from the upper- to the lower-left quadrant of the $E$-plane. 
Although we subtracted the bound-state contribution to $R(E)$, this 
contribution is analytic, having only poles on the negative 
real $E$-axis. Hence, we have analytically continued $R(E)$ from the 
upper- to the lower-left quadrant of the $E$-plane. 

To deal with all values of $E$ in the lower 
half of the complex $E$-plane we introduce ${\bar R}(E)
\equiv \langle {\bar \psi} |G(E)|{\bar \psi} \rangle$, where we 
recall that $|{\bar \psi} \rangle=K|\psi \rangle$. For $\mbox{Im}\; E <0$ 
we can represent ${\bar R}(E)$ by the integral 
\begin{equation}
{\bar R}(E) = -it_0\int_0^{-\infty}d\tau \; e^{i(t_0E)\tau}
{\bar C}(t). \label{eq:403}
\end{equation}
Since ${\bar C(t)}$ is bounded throughout the sector $\pi -\Xi_0 
\leq  \mbox{arg}\; (\tau) < \pi$ we can rotate the contour of 
integration into this sector, and again extend the validity of 
the integral representation, now to a sector in the upper-right 
quadrant of the $E$-plane. To analytically continue ${\bar R}(E)$ 
into the upper-left quadrant of the $E$-plane we proceed as for 
$R(E)$; we project out the bound states, choose $E$ to lie temporarily 
on the left edge of the negative imaginary $E$-axis, rotate the integration 
contour through the lower left quadrant of the $\tau$-plane, 
so that it runs along the left edge of the negative imaginary 
$\tau$-axis, and  subsequently move $E$ into the upper-left quadrant.

Since $C(t)$ and ${\bar C(t)}$ are continuous in the lower-half 
$t$-plane, the integral representations of $R(E)$ and ${\bar R}(E)$ 
are identical when both contours are chosen to be the negative imaginary 
$\tau$-axis. Hence $R(E)$ and ${\bar R}(E)$ are identical in the 
lower-left quadrant of the $E$-plane; it follows that ${\bar R}(E)$ is the 
analytic continuation of $R(E)$ from the upper part of the $E$-plane 
to the lower part, on the first sheet 
of the Riemann energy surface cut from 0 to $\infty$ along a line parallel 
to the integration contour of either $R(E)$ or ${\bar R}(E)$. At the cut, 
$R(E)$ and ${\bar R}(E)$ are different branches of the same function. As 
noted above, the integration contours of $R(E)$ and ${\bar R}(E)$, 
respectively, can be rotated into sectors of the lower- and upper-right 
quadrants of the $E$-plane; hence the cut can be rotated into these 
sectors, thereby extending the region of the Riemann energy surface 
covered by the integral representations. However, these integral 
representations 
do not permit the entire Riemann surface to be covered. As we see below, 
the entire surface can be covered by the series representation. The wider 
region of analyticity afforded by the series representation is a reward 
for performing the integration over time.

Using the expansion of $C(t)$ in powers of $u$ we have
\begin{equation}
R(E) = -it_0\int_0^{\infty}d\tau \; \sum_{n=0}^{\infty}\; 
c_n(t_{\phi})u^n e^{i(Et_0)\tau}. \label{eq:421}
\end{equation}
Rotating the integration contour to the line arg~$(\tau)=\phi$, 
interchanging sum and integral, and recalling that 
\begin{equation}
{\cal I}_n(a) \equiv \int_0^{\infty}d\tau \; e^{ia\tau}
\left( \frac{\tau+i}{\tau-i}\right)^n, \label{eq:422}
\end{equation}
we arrive at the series representation 
\begin{equation}
R(E) = -it_{\phi}
\sum_{n=0}^{\infty}\; c_n(t_{\phi}) {\cal I}_n(Et_{\phi}), \label{eq:426}
\end{equation}
where for the moment $|\phi|<\pi/2$. We obtain the same series representation 
of ${\bar R}(E)$ after expanding ${\bar C}(t)$ in powers of $u$, and 
rotating the integration contour to the line arg~$(\tau)=\pi+\phi$.  
The resolvent itself has the series representation 
\begin{eqnarray}
P_{\rm cont}G(E) & = & -it_0P_{\rm cont}\int_0^{\infty}d\tau \; 
e^{i(Et_0)\tau} U(t) \label{eq:420}\\
= && -it_{\phi}\sum_{n=0}^{\infty}\; {\cal I}_n(Et_{\phi})
P_{\rm cont}C_n(-t_{\phi}H), \label{eq:423}
\end{eqnarray}
as discussed in the Introduction. Since $R(E)$ is defined by an 
integral whose contour is the positive real $\tau$-axis, this 
contour must be mapped into the unit circle, as it is if $\phi$ 
lies in the range $0< \phi < \pi/2$; hence the series representation 
of $R(E)$ is defined for $\phi$ in this range. On the other 
hand, ${\bar R}(E)$ is defined by an integral whose contour is the 
negative real $\tau$-axis, which is mapped into the unit circle if 
$\phi$ lies in the range $-\pi/2< \phi <0$; hence the series representation 
of ${\bar R}(E)$ is defined for $\phi$ in this range. 
It follows that when $E$ is real and 
positive, the physical branch is specified by $0<\phi<\pi/2$, 
whereas the unphysical branch is specified by $-\pi/2<\phi<0$. 

We now prove the useful result that we can pass (discontinuously) from 
one branch of $R(E)$ to the other branch, i.e. 
to ${\bar R}(E)$, by simply changing the sign of the phase $\phi$.
We first observe that, according to Eq.~(\ref{eq:302}), changing the 
sign of $\phi$ and complex conjugating $u$ amounts to changing $\tau$ into 
$-\tau^*$, or, equivalently, $t$ into $-t^*$. It follows from 
Eqs.~(\ref{eq:308}) and (\ref{eq:309c}) that complex conjugating $C(t)$ and 
changing the sign of $\phi$ yields $C(-t^*)$, which, using  
Eq.~(\ref{eq:271}), is the same as $[{\bar C}(t)]^*$. Hence, 
simply changing the sign of $\phi$, without complex conjugation, 
turns $C(t)$ into ${\bar C}(t)$. Furthermore, changing the sign 
of $\phi$, from positive to negative, say, transforms the contour of 
integration for $R(E)$ discontinuously into the contour for ${\bar R}(E)$; 
if we vary $\phi$ continuously from a positive to a negative value 
the integration contour --- the broken line in Fig.\ 3 --- starts as 
a path in the upper semicircle, it moves upwards towards to the 
boundary of this semicircle, and upon reaching this boundary jumps 
discontinuously to the boundary of the lower semicircle, and from 
there to a path inside the lower semicircle. Thus we have arrived at 
the result we set out to prove.


\begin{figure}
  \begin{center}
 \centerline{\psfig{file=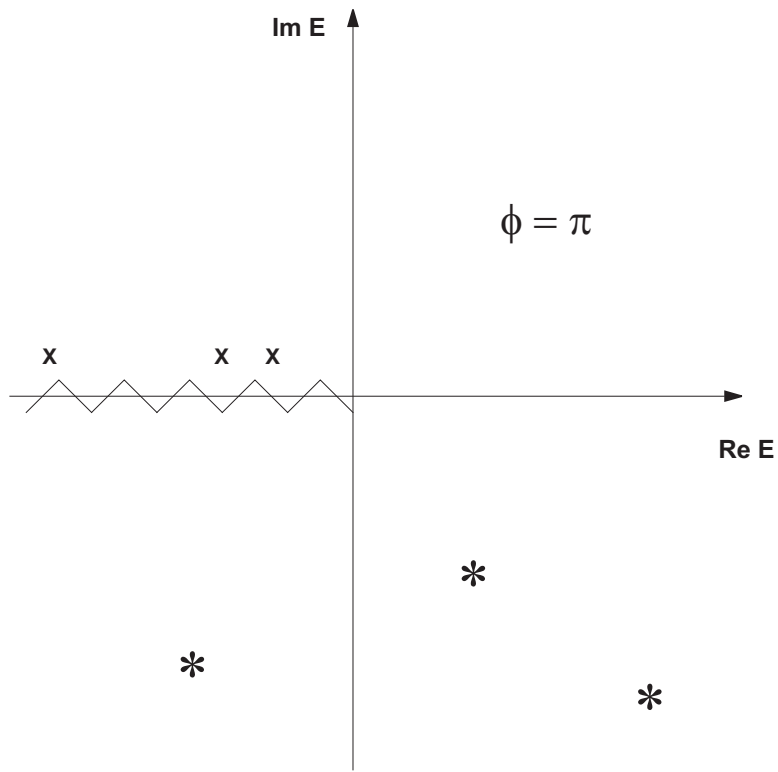,width=4in,height=4in}}
  \end{center}
\caption{The physical sheet of the Riemann energy surface cut along 
the negative real axis. Bound state poles ($\times$) lie on the upper 
edge of the cut and resonance poles ($\star$) lie in the lower-half plane. 
The series representation of the resolvent applies on this sheet 
if we choose the phase parameter to be $\phi=\pi$.\label{fig4}}
\end{figure}


An important task that remains is the analytic continuation 
of the series representation of $R(E)$ from $|\phi|<\pi/2$ to 
$|\phi|>\pi/2$. To carry out this task we must 
analyse the integrals ${\cal I}_n(a)$, defined by 
Eq.~(\ref{eq:422}). These integrals 
are not formally defined when Im~$a\leq 0$ since the integrands 
do not vanish for $\tau \sim \infty$. We can analytically 
continue ${\cal I}_n(a)$ to the region Im~$a\leq 0$ by rotating 
the integration contour into the upper-half $\tau$-plane. 
However, if $n\geq 1$ the integrand is singular along the 
positive imaginary $\tau$-axis, and therefore the (positive) 
angle through which we rotate the contour must be less than $\pi/2$, 
which only permits analytic continuation of ${\cal I}_n(a)$ to the lower-right 
quadrant of the complex $a$-plane. To analytically ${\cal I}_n(a)$ 
continue to the lower-left quadrant of the $a$-plane we move $a$ 
temporarily to the upper edge of the negative real axis in 
the $a$-plane, rotate the integration contour through the angle 
$-\pi$, so that it runs along the lower edge of the negative 
imaginary $\tau$-axis, and subsequently move $a$ into the lower-half 
of the $a$-plane. However, when $n\geq 1$, ${\cal I}_n(a)$ has a branch 
cut extending from 0 to $\infty$ in the $a$-plane since 
we cannot move $a$ continuously around a closed loop by continuously 
rotating the contour of the integral representation of ${\cal I}_n(a)$. 
The direction of this cut is a matter of convention, 
and that convention has already been established. Inspection of 
Eq.~(\ref{eq:423}) reveals that the choice of the cut for 
${\cal I}_n(a)$  determines the cut in the $E$-plane for the 
series representation of $R(E)$, and we want that to be the same 
as for the integral representation of $R(E)$. Thus we draw the cut for 
${\cal I}_n(a)$ along the positive real axis in the $a$-plane, which implies, 
according to Eq.~(\ref{eq:423}), that the series representation 
of $R(E)$ has a cut in the $E$-plane along the line arg~$E=-\phi$.  
This conforms to our specification that $R(E)$ and 
${\bar R}(E)$, respectively, are analytic in the upper and 
lower halves of the $E$-plane, and have integral representations 
defined in the ranges $0<\phi<\pi/2$  and $-\pi/2<\phi<0$. 
We now see how the series can represent $R(E)$ over the entire 
Riemann energy surface. If we let $\phi$ increase from 0 to 
$\pi_-$, the branch cut swings from the positive real $E$-axis, down 
past the resonance poles in the lower-half E-plane, to the lower edge 
of the negative real $E$-axis. Hence, fixing $\phi$ to be $\pi_-$, we 
see that $R(E)$ is defined over an entire energy plane, and its 
poles are the  bound-state poles on the upper edge of the cut and the 
resonance poles in the lower half of the energy-plane. See Fig.\ 4. 
This energy plane is the physical sheet of the Riemann surface, since 
$R(E)$ is the branch corresponding to outgoing-wave boundary conditions when 
$E$ is real and positive. Similarly, if we let 
$\phi$ derease from 0 to $-\pi_-$, the branch cut swings up 
past the resonance poles in the upper-half $E$-plane, to 
the upper edge of the negative real $E$-axis. Hence, fixing 
$\phi$ to be $-\pi_-$, we see that ${\bar R}(E)$ is also defined over 
an entire energy plane, and it has bound-state poles on the lower edge 
of the cut; but its resonance poles are conjugate to those of 
$R(E)$, as shown in Fig.\ 5, and the energy plane is the unphysical sheet 
of the Riemann surface since ${\bar R}(E)$ is the branch corresponding 
to ingoing-wave boundary conditions when $E$ is real and positive. 


\begin{figure}
  \begin{center}
 \centerline{\psfig{file=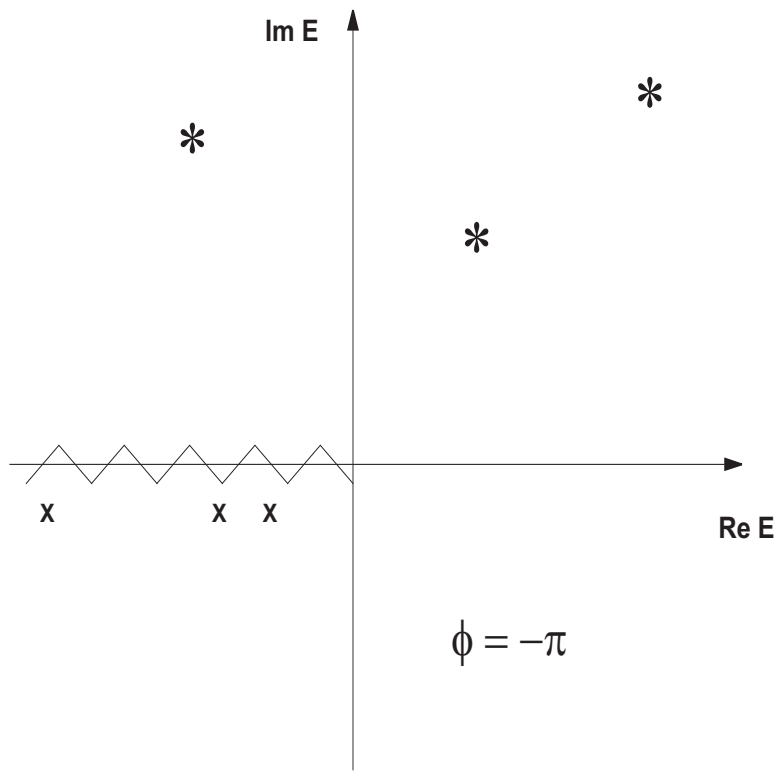,width=4in,height=4in}}
  \end{center}
\caption{The unphysical sheet of the Riemann energy surface cut along 
the negative real axis. Bound state poles ($\times$) lie on the lower 
edge of the cut and resonance poles ($\star$) lie in the upper-half plane. 
The series representation of the resolvent applies on this sheet 
if we choose the phase parameter to be $\phi=-\pi$.\label{fig5}}
\end{figure}


To determine the rate of convergence of the series representation 
of $R(E)$ we need to analyse the behavior of ${\cal I}_n(a)$ for large $n$.
The integrand on the right side of Eq.~(\ref{eq:422}) oscillates 
rapidly when either $n$ or $a$ is large. To treat large $n$ 
it is useful to change to 
the variable $\theta=\tan^{-1}(\tau)$. We have
\begin{equation}
{\cal I}_n(a) =(-1)^n \int_0^{\pi/2}d\theta \; \sec^2 \theta 
e^{ia \tan \theta -2in \theta}. \label{eq:431}
\end{equation}
We start by assuming that $a$ is real and positive and 
that $n>a/2$. There is a point of stationary phase at 
$\theta=\theta_0$, where
\begin{equation}
\cos^2 \theta_0=a/(2n), \label{eq:432}
\end{equation}
and if $n>>a/2$ we have, using the method of stationary phase,
\begin{equation}
{\cal I}_n(a) \sim (-1)^n \sec^2 \theta_0 \left( \frac{2\pi}
{n|s^{''}(\theta_0)|}\right)^{1/2}
e^{ins(\theta_0)+i\pi/4}, \label{eq:433}
\end{equation}
where $s^{''}(\theta)$ is the second derivative with respect to 
$\theta$ of 
\begin{equation}
s(\theta)=(a/n)\tan \theta -2\theta. \label{eq:434}
\end{equation}
Noting that $s^{''}(\theta_0)=4\tan \theta_0$ and that 
\[
\tan \theta_0\approx (2n/a)^{1/2}(1-a/4n) 
\]
we find that
\begin{equation}
{\cal I}_n(a) \sim \left(\frac{2\pi^2n}{a^3}\right)^{1/4}
e^{i\sqrt{8an}+i\pi/4},\; \; \; a> 0.\label{eq:435}
\end{equation}
Evidently, when $a$ is real and positive ${\cal I}_n(a)$ grows as 
$n^{1/4}$ with increasing $n$. When $a$ is real and negative, 
Re~${\cal I}_n(a)=0$; this follows upon rotating the 
contour of integration on the right-side of Eq.~(\ref{eq:422}) 
from the real positive axis to the negative imaginary axis. 
For complex values of $a$, in the plane cut along the positive real 
axis, the point of stationary phase becomes a saddle point, and the 
expression on the right-side of Eq.~(\ref{eq:435}), when 
analytically continued, becomes exponentially small, i.e. it 
decreases exponentially with increasing $n$ as $e^{-\mbox{Im}\; \sqrt{8an}}$. 
However, we must include not just the contribution from the  
saddle point, but also the ``surface'' contribution,  
at the lower limit of the integral ${\cal I}_n(a)$; integration by parts 
yields the following asymptotic expansion in powers of $1/n$, applicable 
as long as the saddle-point contribution is negligible:
\begin{equation}
{\cal I}_n(a)\sim \frac{(-1)^{n+1}i}{n}\sum_{j=0}^{\infty}\; 
\frac{f_j(a)}{n^j} ,\label{eq:435b}
\end{equation}
where $f_j(a)$ is a polynomial of degree $j$ with real 
coefficients. For example, $f_0(a)=1/2$, $f_1(a)=a/4$, and 
$f_2(a)=(a^2-2)/8$. 

Suppose that $E$ is real and positive. Let us set $\phi=0_{\pm}$ 
in the series representation of $R(E)$. The large-$n$ behavior 
of the factor ${\cal I}_n(Et_{\phi})$ is given by 
Eq.~(\ref{eq:435}). These factors, while oscillating as $n$ increases, 
also weakly magnify each term of the series for $R(E)$. 
However, since the $c_n(t_{\phi})$ 
decrease more rapidly than the ${\cal I}_n(Et_{\phi})$ increase, 
the series converges, albeit logarithmically. 
We are now at the stage where we can work through some examples 
which support the validity of the series representations of 
$G(E)$ and $R(E)$, i.e. Eqs.~(\ref{eq:423}) and (\ref{eq:426}). 
We do this in Appendix B. 

If $\phi$ is chosen to be in the range $0<|\phi|<\pi$, the 
saddle-point contribution to ${\cal I}_n(t_{\phi}E)$ can be 
neglected for $n$ sufficiently large, i.e. for 
$\sqrt{8nt_0 E}\sin (|\phi/2|)\gg 1$. In this case we can use 
the asymptotic expansion of Eq.~(\ref{eq:435b}) to express, in a 
rather useful form, the contribution of the higher terms to the 
series representation of $G(E)$. Combining 
Eqs.~(\ref{eq:423}), (\ref{eq:312}), and (\ref{eq:317}) gives
\begin{equation}
P_{\rm cont}G(E) =P_{\rm cont}e^{-t_{\phi}H}\left(\frac{1}{E}
 +it_{\phi}(2t_{\phi}H)S(E,\phi)\right), \label{eq:435c}
\end{equation}
where $S(E,\phi)$ is the infinite sum
\begin{equation}
S(E,\phi)=\sum_{n=1}^{\infty}\; {\cal I}_n(t_{\phi}E)\;_1F_1
(1-n,2,-2t_{\phi}H). \label{eq:435d}
\end{equation}
Using Eqs.~(\ref{eq:435b}) and (\ref{eq:435d}), and  
neglecting the contribution from the saddle point, gives, after 
interchanging the order of the sums,
\begin{equation}
S(E,\phi)\sim -i
\sum_{j=1}^{\infty}\; f_{j-1}(t_{\phi}E)\sum_{n=1}^{\infty}\; 
\frac{(-1)^n}{n^j}\;_1F_1(1-n,2,-2t_{\phi}H). \label{eq:435e}
\end{equation}
We now use the Watson transform:
\begin{equation}
\sum_{n=1}^{\infty}\; (-1)^nF(n)=\frac{1}{2i}\int_{\Gamma} dn \;
\frac{F(n)}{\sin(n\pi)}, \label{eq:435f}
\end{equation}
where $\Gamma$ is a counterclockwise contour which encloses the section of the 
real positive $n$-axis extending from a point in the interval (0,1) to 
infinity. We distort $\Gamma$ into a line, $\Gamma'$, parallel to 
the imaginary $n$-axis, with $0<\mbox{Re}~n<1$, a region 
where it is permissible to use the integral representation
\begin{equation}
_1F_1(1-n,2,z) = \frac{\sin(n\pi)}{n\pi}\int_0^{1}du\; 
e^{zu}\left(\frac{1-u}{u}\right)^n . \label{eq:435g}
\end{equation}
It follows that
\begin{equation}
S(E,\phi)\sim -\sum_{j=1}^{\infty}\; \frac{f_{j-1}(t_{\phi}E)}{2\pi}
\int_{\Gamma} dn \; \frac{1}{n^{j+1}}\int_0^{1}du\; 
e^{-2t_{\phi}Hu}\left(\frac{1-u}{u}\right)^n . \label{eq:435h}
\end{equation}
Define $\omega \equiv \ln[(1-u)/u]$. If $u<1/2$ we have 
$\omega>0$, while if $u>1/2$ we have $\omega<0$. Now, if 
$u>1/2$ we can bend the contour $\Gamma'$ back into 
$\Gamma$; we obtain zero since $\Gamma$ no longer 
encloses any poles. If $u<1/2$ we can bend $\Gamma'$ into 
a contour $\Gamma''$ that encloses the whole of the negative real 
$n$-axis and, in addition, a section $0\leq n \leq n_0 <1$ of the 
positive real $n$-axis. We use, integrating by parts in the first step,
\begin{eqnarray}
\int_{\Gamma''} dn \; \frac{e^{\omega n}}{n^{j+1}} 
 & = & \frac{\omega^j}{j!} \int_{\Gamma''} dn \; \frac{e^{\omega n}}{n} 
\label{eq:435i} \\
& = & \frac{\omega^j}{j!} \left(\int_{-\infty}^{n_0+i\eta} dn \; 
\frac{e^{\omega n}}{n} +\int_{n_0-i\eta}^{-\infty}dn \; 
\frac{e^{\omega n}}{n} \right), \label{eq:435j}
\end{eqnarray}
where $\eta$ is positive but infinitesimal. Transforming the 
integration variable from $n$ to $-n/\omega$, and noting that  
the exponential integral, $E_1(z)$, has a logarithmic 
branch point at $z=0$, with a branch cut customarily drawn along 
the negative real $z$-axis, we obtain
\begin{eqnarray}
\int_{\Gamma''} dn \; \frac{e^{\omega n}}{n} 
 & = & E_1(-\omega n_0+i\eta)-E_1(-\omega n_0-i\eta)
\label{eq:435k} \\
 & = & -2i\pi. \label{eq:435l}
\end{eqnarray}
Putting all this together, we obtain
\begin{equation}
S(E,\phi)\sim i\sum_{j=1}^{\infty}\; \frac{f_{j-1}(t_{\phi}E)}{j!}
\int_0^{1/2}du\; e^{-2t_{\phi}Hu}\left[\ln\left(\frac{1-u}{u}\right)
\right]^j . \label{eq:435n}
\end{equation}
The evolution operator $e^{-2t_{\phi}Hu}$ acts for at most 
one unit of time, and can be calculated rapidly 
(using a Pad\'{e} approximant --- see Appendix C). The first 
few terms --- say the first $N$ --- on the right side of 
Eq.~(\ref{eq:435d}) should be evaluated exactly, and 
accordingly the first $N$ terms built from the right side of 
Eq.~(\ref{eq:435b}) should be subtracted from 
$S(E,\phi)$, with $N$ chosen so that for $n>N$ the 
following conditions are satisfied: (i) The 
asymptotic expansion of Eq.~(\ref{eq:435b}) converges 
when only a small number of terms are included, and (ii) the 
saddle-point contribution to ${\cal I}_n(t_{\phi}E)$ can be 
neglected. \\

{\large\bf 5 Conclusion}\\

We have derived a series representation of the resolvent $G(E)$ 
which incorporates a complex unit of time, $t_0e^{i\phi}$, 
whose phase $\phi$ specifies the branch. This permits 
a real basis to be used to construct a matrix representation 
of the Hamiltonian. The foundation of our approach is the 
analytic continuation of the temporal correlation function 
in the complex time-plane. Using analytic continuation we can 
extrapolate the correlation function from a relatively small 
interval, of duration $t_0$, to an interval of asymptotically 
large duration. Since much of the dynamics take place during 
a time that, typically, is comparable to $t_0$, it should be possible 
to employ a basis efficiently; it is unnecessary to span 
spatial regions that are reached by the particle(s) at times very 
much larger than $t_0$, where the wavefunction takes on its 
asymptotic form. \\[0.5in]

This work was supported by the NSF, Grant No. PHY-9722048.\\

{\large\bf Appendix A: Equivalence of Two Forms of the Integral 
${\cal I}_n(a)$}\\

In this appendix we show that 
Eqs.~(\ref{eq:122}) and (\ref{eq:126}) are equivalent 
definitions of ${\cal I}_n(a)$. To do this we use a standard contour integral 
representation of the Laguerre polynomial:
\begin{equation}
L_{n-1}^{(1)}(x)=\frac{1}{2\pi i}\oint e^{-x\zeta}d\zeta \; 
\left(\frac{1+\zeta}{\zeta}\right)^n , \label{eq:127}
\end{equation}
where the contour runs counterclockwise and encloses the point 
$\zeta=0$. Substituting this 
representation into the right side of Eq.~(\ref{eq:126}) gives
\begin{equation}
{\cal I}_n(Et_{\phi}) = -\frac{1}{\pi}\oint d\zeta \; 
\left(\frac{1+\zeta}{\zeta}\right)^n \int_0^{\infty}dH \; 
e^{-t_{\phi}H(1+2\zeta)}\left(\frac{1}{E-H}-\frac{e^{-t_{\phi}H}}{E}\right). 
\label{eq:128}
\end{equation}
The integrals of Eq.~(\ref{eq:128}) are defined as long as 
(i) $E$ is not real and positive and (ii) Re~$t_{\phi}(1+2\zeta)>0$. 
This second condition implies that the closed contour excludes 
the point $\zeta=-1/2$. The first condition can be dropped after 
integrating over $H$. Performing the integration over $H$ gives, 
with $a=Et_{\phi}$,
\begin{equation}
{\cal I}_n(a) = \frac{1}{2\pi}\oint d\zeta \; 
\left(\frac{1+\zeta}{\zeta}\right)^n \left(
e^{-a(1+2\zeta)}E_1[-a(1+2\zeta)]+\frac{1}{2a(1+\zeta)}\right), 
\label{eq:129}
\end{equation}
where the branch cut of $E_1(z)$, the exponential integral, 
is along the negative real $z$-axis; the contour of integration in 
Eq.~(\ref{eq:129}) excludes the branch point at $\zeta=-1/2$ 
and is drawn to exclude the cut along the line arg~$a(1+2\zeta)=0$. 
Changing variables from $\zeta$ to $\tau=1+2\zeta$ we obtain
\begin{equation}
{\cal I}_n(a) = \frac{1}{2\pi}\oint d\tau \; 
\left(\frac{\tau+1}{\tau -1}\right)^n \left(
e^{-a\tau}E_1(-a\tau)+\frac{1}{a(\tau+1)}\right), \label{eq:130}
\end{equation}
where the new (still counterclockwise) contour excludes both the point 
$\tau=0$ and the branch cut along the line arg~$a\tau=0$, but 
includes the point $\tau=1$. We now deform this contour so that it 
wraps around the branch cut and runs along the circle at infinity. 
Since $e^zE_1(z)\sim (1/z)-(1/z^2)$ for $z\sim \infty$ the 
integral over the circle at infinity vanishes. The integral around 
the branch cut can be simplified after noting that as $z$ moves 
from the upper to the lower edge of the negative real $z$-axis, 
$E_1(z)$ increases by $2\pi i$. It follows that
\begin{equation}
{\cal I}_n(a) = i\int_0^{-\infty} d\tau \; e^{-a\tau}
\left(\frac{\tau+1}{\tau -1}\right)^n . \label{eq:131}
\end{equation}
By rotating the integration contour through $90^{\circ}$, from the 
negative real $\tau$-axis to to the negative imaginary $\tau$-axis, 
we reproduce Eq.~(\ref{eq:122}).\\

{\large\bf Appendix B: Analytical Tests}\\

In this appendix we give two examples where we use the series 
representations given by Eqs.~(\ref{eq:423}) and (\ref{eq:426}) 
to establish some known formal results. In our first example we use 
Eq.~(\ref{eq:423}) to derive the standard expression $1/(E-H)$ for $G(E)$ 
when the potential is arbitrary; but we assume that the part of $G(E)$ 
which accounts for propagation far off the energy shell is relatively 
unimportant. We choose both $t_0$ and $\phi$ to be small, such that $Et_0$ is 
significantly smaller than 1 and $|\phi|\ll \pi$. Hence the exponential 
$e^{i\sqrt{8t_{\phi}En}}$ is almost undamped and is slowly oscillating. 
Therefore a large number of terms contribute to the series on the right 
side of Eq.~(\ref{eq:423}) and the sum is determined primarily by large-$n$ terms.  
Consequently, we can substitute the asymptotic forms for $C_n(-t_{\phi}H)$ 
and ${\cal I}_n(Et_{\phi})$ --- see Eqs.~(\ref{eq:339}) 
and (\ref{eq:435}) --- into the series of Eq.~(\ref{eq:423}) to give
\begin{equation}
G(E) \approx -it_{\phi}\left( \frac{H}{E}\right)^{1/4}
\sum_{n\gg 1}^{\infty}\; \left( \frac{2}{t_{\phi}En}\right)^{1/2}
e^{i\sqrt{8t_{\phi}En}+i\pi/4}\cos(\sqrt{8nt_{\phi}H}+\pi/4). \label{eq:436}
\end{equation}
The cosine on the right side of Eq.~(\ref{eq:436}) has two parts, 
one co-oscillating and the other counter-oscillating relative 
to the pre-exponential. Neglecting the co-oscillating part (its 
contribution averages to zero) gives
\begin{equation}
G(E) \approx -i\left(\frac{t_{\phi}}{2E}\right)^{1/2}
\left( \frac{H}{E}\right)^{1/4}\sum_{n\gg 1}^{\infty}\; 
\frac{e^{i\sqrt{8t_{\phi}En}(1-\sqrt{H/E})}}{\sqrt{n}}. \label{eq:437}
\end{equation}
Replacing the sum over $n$ by an integral over $n$, extending the 
lower limit of integration to zero, and changing variables 
to $y=\sqrt{8En}$, we obtain 
\begin{eqnarray}
G(E) & \approx & -i\frac{\sqrt{t_{\phi}}}{2E}
\left( \frac{H}{E}\right)^{1/4}
\int_0^{\infty}dy\; e^{iy\sqrt{t_{\phi}}(1-H/E)} \label{eq:438} \\
&& =\frac{1}{2E}\left( \frac{H}{E}\right)^{1/4}
\frac{1}{(1-\sqrt{H/E})}, \label{eq:439}
\end{eqnarray}
where we used $\int_0^{\infty}dy\; e^{iy}=i$. Recall that 
when $n$ is large, $C_n(-t_{\phi}H)$ contributes to the 
correlation amplitude $C(t)$ at large times. Therefore 
terms with large $n$ in the series representation of $G(E)$ 
contribute primarily to the near-energy-conserving 
part of $G(E)$; terms with small $n$, which we have neglected, 
contribute to the non-energy-conserving part of $G(E)$. 
Hence it is consistent for us to replace the overall factor 
$(H/E)^{1/4}$ by unity, and the term $(1-\sqrt{H/E})$ by 
$(1-H/E)/2$; this yields $G(E)=1/(E-H)$.

Now we discuss our second example; we derive from Eq.~(\ref{eq:426}) 
the threshold behavior of the rate $-2\mbox{Im}\; R(E)$ for $E\sim 0$. 
Again we need focus 
only on large-$n$ terms on the right side of Eq.~(\ref{eq:426}). 
For simplicity we assume that the angular momentum quantum number 
$l$ is zero. Assuming also that the potential is short-range 
(i.e. no Coulomb tail) it follows from Eqs.~(\ref{eq:426}), (\ref{eq:346}), 
and (\ref{eq:435}) that for $E\sim 0$ 
\begin{equation}
R(E) \sim  -i|{\psi}({\bf 0})|^2 
\left(\frac{15(\pi \mu)^{3/2}}{t_{\phi}^{1/2}(2^{17}
t^3_{\phi}E^3)^{1/4}}\right)\sum_{n\gg 1}^{\infty}\; 
\frac{e^{i\sqrt{8t_{\phi}En}+i\pi/4}}{n^{9/4}}. \label{eq:443}
\end{equation}
Replacing the sum over $n$ by an integral over $n$ with 
lower limit $n_0\gg 1$, and changing variables to $y$ where 
$y^2=\sqrt{8En}$, we obtain
\begin{equation}
R(E) \sim  -i\sqrt{E}|{\psi}({\bf 0})|^2 (\pi \mu)^{3/2}
\left(\frac{15e^{i\pi/4}}
{\sqrt{2}t_{\phi}^{5/4}}\right)\int_{y_0}^{\infty}dy\; 
\frac{e^{iy^2\sqrt{t_{\phi}}}}{y^6}, \label{eq:444}
\end{equation}
where $y_0=(8En_0)^{1/4}$. As $E$ approaches zero, so does $y_0$, 
and the integrand becomes highly singular at the lower limit. 
However, before taking the limit $E\rightarrow 0$ we can 
regularize the integral by integrating by parts three times, and by 
discarding the surface terms at the upper and lower limits; 
the surface terms are exponentially small at the upper limit 
and are of order $1/(y_0t_0^{1/4})$ or less at the lower limit, 
and therefore negligible if  $n_0$ is sufficiently large that 
$8En_0t_0>>1$. Note that the exponential $e^{iy^2\sqrt{t_{\phi}}}$ 
of Eq.~(\ref{eq:444}) restricts $y$ to values less than or of 
order $\phi^{-1/2}t_0^{-1/4}$ and hence $1/(y_0t_0^{1/4})$ cannot 
be less than a number of order $\phi^{1/2}$; therefore 
we require $\phi \ll \pi$. After integrating by parts three times 
we can let the lower limit of the integral be zero, and 
we arrive at
\begin{equation}
R(E) \sim  -i\sqrt{E}t_{\phi}^{1/4}|{\psi}({\bf 0})|^2 (\pi \mu)^{3/2}
2^{7/2}e^{i\pi/4}\int_{0}^{\infty}dy\; e^{iy^2\sqrt{t_{\phi}}}. 
\label{eq:445}
\end{equation}
Performing the integration over $y$ yields, for $E\sim 0$, 
the threshold law 
\begin{equation}
-2\mbox{Im}\; R(E)\sim 2^{7/2}|{\psi}({\bf 0})|^2 
\pi^2 \mu^{3/2}\sqrt{E},  \label{eq:446}
\end{equation}
an expression which of course is independent of $t_{\phi}$. 
This is the correct threshold law if $l=0$ and if the potential is 
short-range. Indeed, as a check we can use 
Eqs.~(\ref{eq:209}) and (\ref{eq:401}) to write, for $E\sim 0$, 
\begin{equation}
-2\mbox{Im}\; R(E)\sim 2^{5/2}(\pi \mu)^{3/2}|{\psi}({\bf 0})|^2 
\mbox{Re}\; e^{-i3\pi/4}\int_0^{\infty}dt\; \frac{e^{iEt}}
{(t+\eta)^{3/2}}, \label{eq:447}
\end{equation}
where we have introduced a positive but infinitesimal term $\eta$ in 
the denominator to regularize the integrand at $t=0$. Integrating 
the right side of Eq.~(\ref{eq:447}) by parts (once), 
and evaluating the resulting integral, we reproduce exactly 
the right side of Eq.~(\ref{eq:446}). For $l\neq 0$ we must 
modify Eq.~(\ref{eq:446}) by including a factor proportional 
to $E^l$ on the right side. 

If the potential has a Coulomb tail it follows from 
Eqs.~(\ref{eq:426}), (\ref{eq:350c}), and (\ref{eq:435}) that
\begin{equation}
R(E) \sim |{\tilde {\psi}}({\bf 0})|^2 
\left(\frac{3\pi^{5/2}Z\mu}{a_0(2^{9}
t^3_{\phi}E^3)^{1/4}}\right)\sum_{n\gg 1}^{\infty}\; 
\frac{e^{i\sqrt{8t_{\phi}En}+i\pi/4}}{n^{7/4}}. \label{eq:448}
\end{equation}
Replacing the sum over $n$ by an integral over $y$, where again
$y^2=\sqrt{8En}$, yields
\begin{equation}
R(E) \sim 12\pi^{5/2}|{\tilde {\psi}}({\bf 0})|^2 e^{i\pi/4}
\left(\frac{Z\mu}{a_0t_{\phi}^{3/4}}\right)
\int_{y_0}^{\infty}dy\; \frac{e^{iy^2\sqrt{t_{\phi}}}}{y^4}, 
\label{eq:449a}
\end{equation}
and, after regularizing the integral by integrating by parts twice, 
and subsequently integrating over $y$, we obtain 
\begin{equation}
R(E) \sim  -i8\pi^{3}(Z\mu/a_0)|{\tilde {\psi}}({\bf 0})|^2, \label{eq:449b}
\end{equation}
which gives the energy-independent threshold law
\begin{equation}
-2\mbox{Im}\; R(E)\sim 16\pi^3 (Z\mu/a_0)|{\tilde {\psi}}({\bf 0})|^2 .
\label{eq:449c}
\end{equation}
As a check we can use Eqs.~(\ref{eq:211b}) and (\ref{eq:401}) to 
write, for $E\sim 0$, 
\begin{equation}
-2\mbox{Im}\; R(E)\sim 32\pi^2 (Z\mu/a_0) |{\tilde {\psi}}({\bf 0})|^2 
\mbox{Im}\; \int_0^{\infty}dt\; \frac{e^{iEt}}{t}; \label{eq:449d}
\end{equation}
since $\int_0^{\infty}dt\; \sin(Et)/t=\pi/2$ we reproduce 
Eq.~(\ref{eq:449c}).\\

{\large\bf Appendix C: Algorithms}\\

 In this appendix we describe some algorithms for the numerical 
implementation of the series representations of $G(E)$ 
and $R(E)$. We consider the 
evaluation of (i) $C_n(z)$, with $z$ a linear operator, 
(ii) ${\cal I}_n(a)$ for all complex numbers $a$, and (iii) $e^z$, 
with $z$ a linear operator. 

(i) \underline{The Operator $C_n(z)$}\\
Recall that $C_n(z)\equiv Q_n(z)e^z$, where $Q_n(z)$ is 
a polynomial of degree $n$ in $z$:
\begin{equation}
Q_n(z)=(2z)\;_1F_1(1-n,2,-2z),\; \; n\geq 1, \label{eq:501}
\end{equation}
with $Q_0(z)=1$. Using a standard recurrence relation for 
the confluent hypergeometric function \cite{Abra} we obtain 
the recurrence relation
\begin{equation}
(n+1)Q_{n+1}(z)=2(n+z)Q_n(z)-(n-1)Q_{n-1}(z),\; 
\; n\geq 1, \label{eq:502}
\end{equation}
which can be started using $Q_1(z)=2z$. Alternatively, we can 
formulate a backward recurrence relation for the function $B_m(z)$ of 
Eq.~(\ref{eq:320}) using a standard recurrence 
relation for the Bessel function. \cite{Abra} We find that 
\begin{equation}
B_{m-1}(z)=-\frac{1}{z}[mB_m(z)-2nB_{m+1}(z)],\; \; m\geq 1. \label{eq:503}
\end{equation}
Since $B_m(z)$ decreases exponentially as $m$ increases, this 
backward recurrence relation can be started by putting 
$B_{m+1}(z)=0$ and $B_m(z)=1$ for a large value of $m$, and by 
using the identity \cite{Abra}
\begin{equation}
B_{0}(z)+2\sum_{m=1}^{\infty}\left(\frac{-2n}{z}\right)^mB_{2m}(z)
=1  \label{eq:504}
\end{equation}
to correctly renormalize the functions. After computing the 
$B_m(z)$ we can use Eq.~(\ref{eq:319}) to calculate $C_n(z)$.

(ii) \underline{The Integrals ${\cal I}_n(a)$}\\
To derive a recurrence relation for the integrals ${\cal I}_n(a)$, 
with Im~$a>0$, we start from Eq.~(\ref{eq:431}). Using
\begin{equation}
\cos^2 \theta =(2+e^{2i\theta}+e^{-2i\theta})/4,  \label{eq:505}
\end{equation}
we see that
\begin{equation}
2{\cal I}_n(a)-{\cal I}_{n-1}(a)-{\cal I}_{n+1}(a)=
4(-1)^n\int_0^{\pi/2}d\theta \; 
e^{ia\tan \theta-2in\theta} . \label{eq:506}
\end{equation}
Integrating by parts on the right side of Eq.~(\ref{eq:506}), and 
discarding the surface term at the upper limit $\theta =\pi/2$ 
(it vanishes if Im~$a>0$) yields
\begin{equation}
2{\cal I}_n(a)-{\cal I}_{n-1}(a)-{\cal I}_{n+1}(a)=
(-1)^{n+1}\left(\frac{2i}{n}\right)+
(-1)^n\left(\frac{2a}{n}\right)\int_0^{\pi/2}d\theta \; \sec^2 \theta 
e^{ia\tan \theta-2in\theta} . \label{eq:507}
\end{equation}
Recognizing that the integral on the right side of Eq.~(\ref{eq:507}) 
is proportional to ${\cal I}_n(a)$ we arrive at the recursion formula
\begin{equation}
{\cal I}_{n+1}(a) = (-1)^n\left(\frac{2i}{n}\right)+2\left(
1-\frac{a}{n}\right){\cal I}_{n}(a)-{\cal I}_{n-1}(a), \; \; 
n\geq 1,  \label{eq:508}
\end{equation}
which can be started using
\begin{eqnarray}
{\cal I}_0(a) &=& i/a, \label{eq:509} \\
{\cal I}_1(a) &=& {\cal I}_0(a)+2ie^{-a}\mbox{E}_1(-a), \label{eq:510} 
\end{eqnarray}
where $E_1(z)$ is the exponential integral. As long as 
$a$ is positive (with an infinitesimal positive imaginary part)  
${\cal I}_n(a)$ increases 
(albeit weakly) as $n$ increases, and this forward recurrence relation 
is stable. However, for large $a$ we have (see below) 
${\cal I}_n(a) \sim (-1)^n (i/a)$ and therefore there is cancellation 
between the first and second terms on the right side of 
Eq.~(\ref{eq:508}). Thus we consider separately, below, the 
case $a \gg n$. 

Note that $E_1(z)$ has a logarithmic branch point singularity at $z=0$, 
and this singularity, which first appears in Eq.~(\ref{eq:510}), 
is propagated by the recursion formula of Eq.~(\ref{eq:508}) so 
that all terms but the first on the right-side of Eq.~(\ref{eq:426}) 
have logarthmic branch points at $E=0$ [the branch cuts lie along 
the line arg~$(E)=-\phi$]. In contrast, the exact $R(E)$, which is 
a function of the dimensionless variable $Et_0$, has a square-root 
branch point at $E=0$ (if the potential is short-range). Presumably the 
{\em converged} sum on the right-side of Eq.~(\ref{eq:426}) does have 
the correct branch-point behavior. In this regard it is helpful 
consider an example, namely the function  
$f(z)=\sum_{n=0}^{\infty}\; [\ln(z)/2]^n/n!$. Any approximation 
to $f(z)$ obtained by truncating the sum has a logarithmic branch point; 
but the infinite sum is $f(z)=\exp[\ln(z)/2]=z^{1/2}$, which has 
no logarithmic branch point, only a square root branch point.

If $|a| \gg n$ we can represent ${\cal I}_n(a)$ by an asymptotic power 
series in $1/a$. Integrating by parts on the right side of 
Eq.~(\ref{eq:422}) yields 
\begin{equation}
{\cal I}_n(a) =\sum_{m=0}^{\infty} \left( \frac{i}{a}\right)^{m+1}
Y_n^{(m)}(0), \label{eq:511}
\end{equation}
where $Y_n^{(m)}(\tau)$ is the $m$-th derivative with respect to $\tau$ 
of
\begin{equation}
Y_n^{(0)}(\tau)=\left( \frac{\tau+i}{\tau-i}\right)^n. 
\label{eq:512}
\end{equation}
We have $Y_n^{(0)}(0)=(-1)^n$ and the derivatives can be 
calculated using the following relation (which can be obtained 
after some straightforward algebra)
\begin{equation}
Y_n^{(m+1)}(\tau)=2n\sum_{k=0}^{[m/2]} (-i)^{2k+1}\frac{m!}{(m-2k)!}
Y_n^{(m-2k)}(0), \label{eq:514}
\end{equation}
where $[m/2]$ is $m/2$ rounded to the lowest integer. It 
follows from Eq.~(\ref{eq:514}) that $Y_n^{(m)}(0)$ is a real number 
multiplied by $i^m$, and hence if $a$ is real each term of the 
asymptotic series (\ref{eq:511}) is purely imaginary. In fact, the real 
part of ${\cal I}_n(a)$ is exponentially small; to see this, rotate the 
contour of integration on the right side of Eq.~(\ref{eq:422}) 
by $180^{\circ}$, counterclockwise. This yields ${\cal I}_n(a)^*$, and 
since there is a pole at $\tau=i$ it follows from Cauchy's theorem 
that the real part of ${\cal I}_n(a)$ is $\pi i$ multiplied by the 
residue of this pole. Thereby we arrive at
\begin{eqnarray}
\mbox{Re}\; {\cal I}_n(a) &=& e^{-a}
\left(\frac{\pi}{a}\right)\sum_{m=0}^{n-1} 
\; \frac{n!(-2a)^{n-m}}{m!(n-1-m)!(n-m)!} \label{eq:515a} \\
&& = -2\pi e^{-a}\sum_{k=0}^{n-1} \; \frac{n!}{(k+1)!(n-k-1)!}
\frac{n!(-2a)^k}{k!} \label{eq:515b} \\
&& = -2\pi e^{-a}L_{n-1}^{(1)}(2a),\; \; \; \mbox{Im}\; a=0,
\; \; \; a>0  \label{eq:515c} 
\end{eqnarray}
Note again that if $a$ is real and negative, Re~${\cal I}_n(a)=0$.

(iii) \underline{Exponentiation of a Linear Operator}\\
To evaluate $e^z$ we use the Pad\'{e} approximant 
\begin{equation}
e^z\approx \frac{1+\frac{z}{2}+\frac{z^2}{12}}
{1-\frac{z}{2}+\frac{z^2}{12}}, \label{eq:516} 
\end{equation}
which matches the power series expansion of $e^z$ through 
the term in $z^4$. Since $z$ is an operator it is convenient 
to factorize this expression so that we do not incur the 
additional computation of $z^2$; thus we write
\begin{equation}
e^z\approx \frac{(1-z/z_1)(1-z/z_2)}{(1+z/z_1)(1+z/z_2)}, 
\label{eq:517} 
\end{equation}
where the roots $z_1$ and $z_2$ are
\begin{equation}
z_{1,2}=-3\pm i\sqrt{3}. \label{eq:518} 
\end{equation}
If the Hamiltonian $H$ is represented by the matrix $\underline{H}$, 
constructed from a basis with overlap matrix $\underline{B}$, the 
time-evolution operator $e^{-iHt}$ becomes 
$e^{-i(\underline{B}^{-1}\underline{H})t}$, and we have 
(letting $t\rightarrow -it$) 
\begin{equation}
e^{-Ht} \approx 
\frac{1}{\underline{B}-(t/z_1)\underline{H}}
[\underline{B}+(t/z_1)\underline{H}]
\frac{1}{\underline{B}-(t/z_2)\underline{H}}
[\underline{B}+(t/z_2)\underline{H}]. \label{eq:519} 
\end{equation}
This is an extension of the standard Cayley form of the time-evolution 
operator from third- to fifth-order; we have found the stability 
and accuracy of exponentiation to be significantly improved by 
this increase in order. While it is unnecessary to calculate 
$\underline{B}^{-1}$ here, note that since $\underline{B}$ is real, 
symmetric, and positive definite it has a Cholesky 
decomposition \cite{NumR} and hence its inverse can be calculated rapidly.\\

{\large\bf Appendix D: Large-$n$ form of $c_n(t_{\phi})$ when Coulomb 
Tail is Present}\\

To establish Eqs.~(\ref{eq:350c}) and Eqs.~(\ref{eq:350d}) for 
the large-$n$ form of $c_n(t_{\phi})$ when a Coulomb tail is 
present, we first show how to analytically continue $c_{{\rm bd},n}(t_{\phi})$ 
along a path in the upper-half $t_{\phi}$-plane,  
from the upper edge of the cut 
to the positive real axis on the first sheet of the Riemann 
$t_{\phi}$-surface. To this end we seek to 
replace the sum over $m$ on the right side of Eq.~(\ref{eq:349d}) 
by an integral over $m$. However, the direct replacement of sum 
by integral is not entirely justified 
since the summand varies rapidly when $n\sim \infty$, 
and the $m$-th and $(m+1)$-th terms of the sum 
may differ significantly. Nevertheless, 
we can accomplish the passage from sum to 
integral by making a small modification to the integrand. 
We first break the cosine in the summand into a sum of 
two exponentials:
\begin{equation}
c_{{\rm bd},n}(t_{\phi}) \sim 8\pi^2 Z^3 |{\tilde \psi}({\bf 0})|^2 
\left(\frac{2}{\pi^2 n^3}\right)^{1/4}[S_{+}(t_{\phi})+S_{-}(t_{\phi})],
\label{eq:349p} 
\end{equation}
where
\begin{equation}
S_{\pm}(t_{\phi})]= \sum_{m=1}^{\infty} 
\frac{(t_{\phi}E_{{\rm bd},m})^{1/4}}{2(m^*a_0)^3}
e^{\pm i\sqrt{8nt_{\phi}E_{{\rm bd},m}}+\pm i\pi/4}, 
\label{eq:349q} 
\end{equation}
and we introduce the integrals 
\begin{equation}
I_{+}(t_{\phi}) \equiv  \int_{C_{+}}dm\;  
\frac{(t_{\phi}E_{{\rm bd},m})^{1/4}}{2(m^*a_0)^3}
e^{i\sqrt{8nt_{\phi}E_{{\rm bd},m}}+i\pi/4}
\label{eq:349r} 
\end{equation}
and 
\begin{equation}
I_{-}(t_{\phi}) \equiv  \int_{C_{-}}dm\;  
\frac{(t_{\phi}E_{{\rm bd},m})^{1/4}}{2(m^*a_0)^3}
\frac{e^{-i\sqrt{8nt_{\phi}E_{{\rm bd},m}}-i\pi/4}}{(1-e^{2i\pi m})}, 
\label{eq:349s} 
\end{equation}
where the contours $C_{\pm}$ run from $m=1$ to $\infty$ along the 
upper edge of the positive real $m$-axis (and where $m^*$ 
is now the continuous variable $m-\delta$). We have 
introduced the factor $(1-e^{2i\pi m})$ into the denominator 
of the integrand of $I_{-}(t_{\phi})$, a factor which 
vanishes when $m$ is an integer. The integral $I_{+}(t_{\phi})$ is 
formally defined for all $t_{\phi}$ on the first sheet, except on 
the cut, of the Riemann  $t_{\phi}$-surface, i.e. for 
$-\pi < \phi <\pi$, since $i\sqrt{8nt_{\phi}E_{{\rm bd},m}}
=-\sqrt{8nt_{\phi}|E_{{\rm bd},m}|}$ and the exponential in the 
integrand of $I_{+}(t_{\phi})$ decays as $m$ increases. 
Furthermore, due to this exponential, only the region in which 
$|8nt_{\phi}E_{{\rm bd},m}|$ is less than or of order unity contributes 
significantly to $I_{+}(t_{\phi})$; in this region $m^*$, and hence $m$, 
are greater than or of the order of $\sqrt{n}$, and so the integrand 
varies slowly with $m$. Consequently, when $t_{\phi}$ is on the 
upper edge of the cut the integral $I_{+}(t_{\phi})$ and the sum 
$S_{+}(t_{\phi})$ are the same. The integral $I_{-}(t_{\phi})$ is 
formally defined for all $t_{\phi}$ on the second sheet reached by 
crossing the branch cut, i.e. for $\pi < \phi < 3\pi$. 
For $t_{\phi}$ on the branch cut, the exponential in the 
numerator of the integrand of $I_{-}(t_{\phi})$ 
oscillates, but it is undamped so we cannot directly replace the integral 
by the sum. However, we can express the integral as a sum 
by observing that the integrand of $I_{-}(t_{\phi})$ has an infinite 
sequence of poles on the positive real $m$-axis, on the lower edge 
of the contour $C_-$, at those points where 
$m$ is a nonnegative integer. If we rotate $C_-$ downwards, past 
the line of poles, we pick up the contributions from the poles according 
to Cauchy's residue theorem. The integral along the new, rotated, 
contour is negligible for the following reason: The exponential in the 
numerator of the integrand is now damped, and so the main contribution 
to the integral comes from the region where $m$ is greater 
than or of the order of $\sqrt{n}$; but the exponential 
$e^{2i\pi m}$ in the denominator is exponentially large since 
Im~$(m)$ is large and negative along the important segment of the 
rotated contour. It follows that for $t_{\phi}$ on the branch cut, 
and anywhere on the second sheet, $I_{-}(t_{\phi})$ is, when $n\sim \infty$, 
just the sum of the residues of the poles multiplied by $-2\pi i$; this 
is the same as $S_{-}(t_{\phi})$ when $t_{\phi}$ is on the cut.

To evaluate $I_{+}(t_{\phi})$ we change variables from $m$ to 
$y=\sqrt{Ze/(m^*a_0^{1/2})}$. The contour of integration runs 
from $y=0$ to roughly $\sqrt{Ze/(a_0^{1/2})}$. 
Extending the upper limit to $\infty$ (which does not affect 
the leading $n$-dependence of the integral) and rotating the 
contour of integration downwards through an angle of $\phi/4$, 
replacing $y$ by $ye^{-i\phi/4}$, gives
\begin{eqnarray}
I_{+}(t_{\phi})& =& i\left(\frac{e^{-i\phi}\mu}{Z^2a_0}\right)
\left(\frac{t_0}{2}\right)^{1/4}
\int_0^{\infty} dy\; y^4e^{-2y^2\sqrt{nt_0}} \nonumber \\
&=&\frac{i}{n^{5/4}}\left(\frac{3\pi^{1/2}\mu}{2^{23/4}Z^2a_0t_{\phi}}
\right),  \; \; \; -\pi < \phi <\pi .  \label{eq:349t} 
\end{eqnarray}
A similar change of variables for $I_{-}(t_{\phi})$ gives, after 
extending the upper limit of integration to $\infty$ and rotating the 
contour of integration upwards through an angle of $\Theta$, 
\begin{equation}
I_{-}(t_{\phi}) =e^{i5\Theta}\left(\frac{e^{i\phi/4}\mu}{Z^2a_0}\right)
\left(\frac{t_0}{2}\right)^{1/4}
\int_0^{\infty} dy\; y^4\frac{e^{2y^2e^{i2(\Theta+\phi/4)}
\sqrt{nt_0}}}{(1-e^{2i\pi \delta}e^{2i\pi (e^{-2i\Theta}Ze/a_0^{1/2}y^2)})}.
\label{eq:349u} 
\end{equation}
The integral $I_{-}(t_{\phi})$ is formally defined if 
$(3\pi-\phi)/4>\Theta >(\pi-\phi)/4$. We set $\Theta$ equal to its 
minimum value $(\pi_+ -\phi)/4$. Let us, for the moment, choose 
$t_{\phi}$ to be on the second sheet, so that 
$\phi>\pi$; this implies that $\Theta<0$. Due to the exponential in the 
numerator on the right side of Eq.~(\ref{eq:349u}), the main 
contribution to the integral comes from values of $y$ less than 
or of order $1/n^{1/4}$, and since $\Theta<0$ the 
exponential $e^{2i\pi (e^{-2i\Theta}Ze/a_0^{1/2}y^2)}$ 
in the denominator on the right side of Eq.~(\ref{eq:349u}) 
is negligible. It follows that 
\begin{eqnarray}
I_{-}(t_{\phi}) &\sim& e^{i5\Theta}\left(\frac{e^{i\phi/4}\mu}{Z^2a_0}\right)
\left(\frac{t_0}{2}\right)^{1/4}
\int_0^{\infty} dy\; y^4 e^{2y^2e^{i2(\Theta+\phi/4)}\sqrt{nt_0}} 
\nonumber \\
&=& -\frac{i}{n^{5/4}}\left(\frac{3\pi^{1/2}\mu}{2^{23/4}Z^2
a_0t_{\phi}}\right),  \; \; \; \pi < \phi < 3\pi. \label{eq:349v} 
\end{eqnarray}
[The right sides of Eqs.~(\ref{eq:349t})  and (\ref{eq:349v}) 
differ only by a sign.] 
Now let us choose $t_{\phi}$ to be above the branch cut on the 
first sheet, so that $\phi<\pi$; this implies $\Theta>0$. Again, 
the main contribution to the integral comes from values of $y$ less than 
or of order $1/n^{1/4}$, but since $\Theta>0$ the 
exponential $e^{2i\pi (e^{-2i\Theta}Ze/a_0^{1/2}y^2)}$ 
is now very large. Hence, as $t_{\phi}$ crosses the cut from below 
to above, moving from the second sheet to the first, 
$I_{-}(t_{\phi})$ vanishes through order $1/n^{5/4}$. It follows 
that for $t_{\phi}$ on the first sheet
\begin{equation}
c_{{\rm bd},n}(t_{\phi})  \sim
\frac{i}{n^2}\left( \frac{3\pi^2Z\mu |{\tilde \psi}({\bf 0})|^2}
{2^{5/2}a_0t_{\phi}}\right),  \; \; \; -\pi < \phi <\pi  ,
\label{eq:349w} 
\end{equation}
but that $c_{{\rm bd},n}(t_{\phi})$ vanishes through order 
$1/n^2$ on the second sheet, i.e. when $\pi < \phi < 3\pi$. For 
$\phi \neq \pm \pi_-$ we must project out those bound states for which 
$m^*$ is roughly less than $\sqrt{n}$.

We now analyse $c_{{\rm cont},n}(t_{\phi})$ when the potential 
has an attractive Coulomb tail. The integral over $k$ on the left side 
of Eq.~(\ref{eq:345}) is, strictly speaking, defined 
only when $\phi=0$, since $\cos(\sqrt{8nt_{\phi}E_k}+\pi/4)$ 
diverges as $k$ increases unless $t_{\phi}$ is real and positive. 
However, as before we break the cosine into a sum of two exponentials.  
After performing the integration, we can analytically continue the 
separate integrals. Let us start with $t_{\phi}=t_0$ and move $t_{\phi}$ 
along a path in the upper-half plane to the point $t_0e^{i\pi_-}$. 
We first change variables from $k$ to $y=(k/\sqrt{\mu})^{1/2}$, which 
removes from the integrand the branch point arising from the 
factor $k^{1/2}$. Note that $i\sqrt{8nt_{\phi}E_k}$ is equal to 
$2ie^{i\phi/2}y^2\sqrt{nt_0}$, and 
has a negative real part over the entire path. It follows that, 
after breaking the cosine into a sum of two exponentials, the 
integral over the term in $\exp (i\sqrt{8nt_{\phi}E_k}+i\pi/4)$ 
is well-defined over the entire path since its integrand 
decreases exponentially as $y^2$ increases. On the other hand, the 
integral over the term in $\exp (-i\sqrt{8nt_{\phi}E_k}-i\pi/4)$ 
is not well-defined; its integrand increases exponentially as 
$y^2$ increases. To analytically continue this second integral 
we proceed as before, and rotate the contour of $y$-integration 
downwards through an angle $\Theta$ where $\Theta>\phi/4$ so that the 
exponential does not explode with increasing $y^2$ as we move 
$t_{\phi}$. The first integral is
\begin{eqnarray}
\frac{1}{2}\int d^3 k\; (t_{\phi}E_k)^{1/4}|{\psi}({\bf k})|^2
e^{i\sqrt{8nt_{\phi}E_k}+i\pi/4} &=& \frac{2\pi Z\mu}{a_0}
\left( \frac{e^{i\pi}t_{\phi}}{2}\right)^{1/4} \nonumber \\
&& \times \int d\Omega \int_0^{\infty} dy\;  
\frac{y^4|{\tilde \psi}({\bf k})|^2}{1-e^{-2\pi \gamma}}
e^{2iy^2\sqrt{t_{\phi}n}}, \label{eq:349a} 
\end{eqnarray}
where $\gamma=Z/(a_0\mu^{1/2}y^2)$ and where $d\Omega$ is an element of 
solid angle containing ${\bf k}$. We now change variables in the first 
integral from $y$ to $x=yn^{1/4}$ so that $k=x^2\sqrt{\mu/n}$ 
and $\gamma=(Z/a_0x^2)\sqrt{n/\mu}$; for $n\sim \infty$ 
we can neglect $e^{-2\pi \gamma}$ compared to 1, and we can 
replace $|{\tilde \psi}({\bf k})|^2$ by 
$|{\tilde \psi}({\bf 0})|^2$, so the first integral becomes
\begin{equation}
\frac{2\pi Z\mu}{a_0}\frac{|{\tilde \psi}({\bf 0})|^2}{n^{5/4}}
\left( \frac{e^{i\pi}t_{\phi}}{2}\right)^{1/4}\int d\Omega 
\int_0^{\infty} dx\;  
x^4 e^{2ix^2\sqrt{t_{\phi}}}= -\frac{i}{n^{5/4}}\left( \frac{3\pi^{5/2}Z\mu 
|{\tilde \psi}({\bf 0})|^2}{2^{11/4}a_0t_{\phi}}\right). 
\label{eq:349c} 
\end{equation}
The second integral is, after changing variables from $y$ to $ye^{-i\Theta}$,
\begin{eqnarray}
\frac{1}{2}\int d^3 k\; (t_{\phi}E_k)^{1/4}|{\psi}({\bf k})|^2
e^{-i\sqrt{8nt_{\phi}E_k}-i\pi/4} &=& \frac{2\pi Z\mu e^{-i5\Theta}}{a_0}
\left( \frac{e^{-i\pi}t_{\phi}}{2}\right)^{1/4} 
\int d\Omega \int_0^{\infty} dy\; \nonumber \\ 
&\times& \frac{y^4|{\tilde \psi}({\bf k})|^2}
{1-e^{-2\pi \gamma}} e^{-2iy^2e^{-2i\Theta}\sqrt{t_{\phi}n}},
\label{eq:349b} 
\end{eqnarray}
where now $k=\mu^{1/2}y^2e^{-2i\Theta}$ and 
$\gamma=Ze^{2i\Theta}/(a_0\mu^{1/2}y^2)$. We fix $\Theta$ to have its 
minimum value, i.e. $\Theta =\phi/4+\eta/2$, with $\eta$ positive 
but infinitesimal and change variables from $y$ to 
$x=yn^{1/4}$. For $n\sim \infty$ we have
\begin{eqnarray}
\frac{1}{2}\int d^3 k\; (t_{\phi}E_k)^{1/4}|{\psi}({\bf k})|^2
e^{-i\sqrt{8nt_{\phi}E_k}-i\pi/4} &\sim& \frac{8\pi^2 Z\mu 
e^{-i5\Theta}}{a_0}\frac{|{\tilde \psi}({\bf 0})|^2}{n^{5/4}}
\left( \frac{e^{-i\pi}t_{\phi}}{2}\right)^{1/4} \nonumber \\
&\times& \int_0^{\infty} dx\;  x^4 
\frac{e^{-2(i+\eta)x^2\sqrt{t_0}}}{1-e^{-2\pi \gamma}}, \label{eq:349i} 
\end{eqnarray}
where $k=x^2e^{-2i\Theta}\sqrt{\mu/n}$ and 
$\gamma=e^{2i\Theta}(Z/a_0x^2)\sqrt{n/\mu}$. Note that the integrand 
has an infinite number of poles, accumulating at $y=0$ on the line 
arg~$(y)=-\pi/4$, at those points where $e^{-2\pi \gamma}=1$. As long as 
$-\pi<\phi<\pi$ we have $-\pi/2 <2\Theta <\pi/2$, so that 
the contour of integration lies apart from the line of poles, and, 
furthermore, Re~$\gamma\gg 1$ so that 
the term in $e^{-2\pi \gamma}$ on the right side of 
Eq.~(\ref{eq:349i}) is negligible; hence we 
can integrate over $x$ to give
\begin{equation}
\frac{1}{2}\int d^3 k\; (t_{\phi}E_k)^{1/4}|{\psi}({\bf k})|^2
e^{-i\sqrt{8nt_{\phi}E_k}-i\pi/4} \sim
\frac{i}{n^{5/4}}\left( \frac{3\pi^{5/2}Z\mu |{\tilde \psi}({\bf 0})|^2}
{2^{11/4}a_0t_{\phi}}\right), \; \; \; -\pi< \phi <\pi. 
\label{eq:349k} 
\end{equation}
Therefore, when $-\pi< \phi<\pi$ the two integrals have 
leading terms that are equal but opposite, and 
$c_{{\rm cont},n}(t_{\phi})$ decreases faster than $n^{-2}$ as 
$n$ increases. However, we now let $\phi$ approach $\pi$. This forces  
$\Theta$ to approach $\pi_+/4$, and the integration contour moves 
across the line of poles, to its lower edge, resulting in a rapid 
change in the integral by an amount equal to the contribution of the 
poles. To determine the value of the new integral we increase 
$\Theta$ further, holding $\phi$ fixed at $\pi_+/4$, so that 
$\Theta$ moves into the range $\pi/2 < 2\Theta < 3\pi/2$ where 
Re~$\gamma$ is large and negative and $e^{-2\pi \gamma}$ is 
exponentially large; it follows that the new integral is negligible. 
Hence the integral on the left side of Eq.~(\ref{eq:349k}) is 
negligible when $t_{\phi}$ lies on the second sheet. It follows 
that while $c_{{\rm cont},n}(t_{\phi})$ vanishes through order 
$1/n^2$ when $t_{\phi}$ lies on the first sheet, 
\begin{equation}
c_{{\rm cont},n}(t_{\phi})  \sim
-\frac{i}{n^2}\left( \frac{3\pi^2Z\mu |{\tilde \psi}({\bf 0})|^2}
{2^{5/2}a_0t_{\phi}}\right),  \; \; \; \pi < \phi <3\pi  ,
\label{eq:349x} 
\end{equation}
when $t_{\phi}$ lies on the second sheet. We have now established 
Eqs.~(\ref{eq:350c}) and Eqs.~(\ref{eq:350d}).

\end{document}